# Extremely well isolated 2D spin-1/2 antiferromagnetic Heisenberg layers with small exchange coupling in the molecular-based magnet CuPOF

D. Opherden,[1, 2, *] N. Nizar,[3] K. Richardson,[3] J. C. Monroe,[4] M. M. Turnbull,[4] M. Polson,[5] S. Vela,[6]

W. J. A. Blackmore,[7] P. A. Goddard,[7] J. Singleton,[8] E. S. Choi,[9] F. Xiao,[10] R. C. Williams,[10]

T. Lancaster,[10] F. L. Pratt,[11] S. J. Blundell,[12] Y. Skourski,[1] M. Uhlarz,[1] A. N. Ponomaryov,[1]

S. A. Zvyagin,[1] J. Wosnitza,[1, 2] M. Baenitz,[13] I. Heinmaa,[14] R. Stern,[14] H. Kühne,[1, †] and C. P. Landee[3, ‡]

[1] *Hochfeld-Magnetlabor Dresden (HLD-EMFL) and Würzburg-Dresden Cluster of Excellence ct.qmat,*
*Helmholtz-Zentrum Dresden-Rossendorf, 01328 Dresden, Germany*
[2] *Institut für Festkörper- und Materialphysik, TU Dresden, 01062 Dresden, Germany*
[3] *Department of Physics, Clark University, Worcester, MA 01610, USA*
[4] *Carlson School of Chemistry, Clark University, Worcester, MA 01610, USA*
[5] *Department of Chemistry, University of Canterbury, Private Bag 4800, Christchurch, New Zealand*
[6] *Laboratoire de Chimie Quantique, UMR 7177, CNRS-Universitéde Strasbourg,*
*4, rue Blaise Pascal, F-67000 Strasbourg, France*
[7] *Department of Physics, University of Warwick, Gibbet Hill Road, Coventry, CV4 7AL, UK*
[8] *National High Magnetic Field Laboratory, Los Alamos National Laboratory, Los Alamos, New Mexico 87545, USA*
[9] *National High Magnetic Field Laboratory, Florida State University, Tallahassee, FL 32310, USA*
[10] *Durham University, Centre for Materials Physics, South Road, Durham DH1 3LE, UK*
[11] *ISIS Facility, Rutherford Appleton Laboratory, Chilton, Didcot, Oxford OX11 0QX, UK*
[12] *Clarendon Laboratory, Department of Physics, University of Oxford, Park Road, Oxford OX1 3PU, UK*
[13] *Max Planck Institute for Chemical Physics of Solids, 01187 Dresden, Germany*
[14] *National Institute of Chemical Physics and Biophysics, Akadeemia tee 23, 12618 Tallinn, Estonia*
(Dated: September 1, 2020)

We report on a comprehensive characterization of the newly synthesized Cu$^{2+}$-based molecular magnet $[Cu(pz)_2(2\text{-}HOpy)_2](PF_6)_2$ (CuPOF), where pz = $C_4H_4N_2$ and 2-HOpy = $C_5H_4NHO$. From a comparison of theoretical modeling to results of bulk magnetometry, specific heat, $\mu^+$SR, ESR, and NMR spectroscopy, this material is determined as an excellent realization of the 2D square-lattice $S = 1/2$ antiferromagnetic Heisenberg model with a moderate intraplane nearest-neighbor exchange coupling of $J/k_B = 6.80(5)$ K, and an extremely small interlayer interaction of about 1 mK. At zero field, the bulk magnetometry reveals a temperature-driven crossover of spin correlations from isotropic to $XY$ type, caused by the presence of a weak intrinsic easy-plane anisotropy. A transition to long-range order, driven by the low-temperature $XY$ anisotropy under the influence of the interlayer coupling, occurs at $T_N = 1.38(2)$ K, as revealed by $\mu^+$SR. In applied magnetic fields, our $^1$H-NMR data reveal a strong increase of the magnetic anisotropy, manifested by a pronounced enhancement of the transition temperature to commensurate long-range order at $T_N = 2.8$ K and 7 T.

## I. INTRODUCTION

The study of critical phenomena, related to phase transitions between exotic ground states that emerge from complex underlying electronic correlations, is a subject of high importance in the research of low-dimensional magnetism. As has been established by extensive theoretical work, in contrast to the cases of one- and three-dimensional magnetic lattices, the critical behavior and ground-state properties of 2D lattice systems are strongly dependent on the symmetry of the interactions between the magnetic moments, e.g., Ising, $XY$, and Heisenberg types [1]. For instance, in accordance with Onsager's exact solution [2], the 2D spin-1/2 Ising antiferromagnet undergoes a Néel-type transition to long-range order at

$T_N = 1.06J/k_B$ [3], where $J$ is the exchange strength between neighboring magnetic moments. In contrast, thermal fluctuations in the 2D quantum Heisenberg antiferromagnet (2D QHAF) prevent long-range order at any finite temperature, as was rigorously proven by Mermin and Wagner [4]. The 2D planar, or $XY$, magnetic lattice does not exhibit Néel-type order, despite the divergence of the susceptibility at finite temperatures. Instead, an unusual topological order with characteristic algebraic decay of the spin correlations was proposed by Berezinskii [5], Kosterlitz and Thouless [6]. Here, the formation of bound pairs of topological excitations, called vortex and antivortex states, characterizes the ordered state, where the unbinding of the vortex-antivortex pairs constitutes the Berezinskii-Kosterlitz-Thouless (BKT) transition, which occurs at $T_{BKT} = 0.353J/k_B$ for the $S = 1/2$ case [7].

Advancing the experimental research of phase transitions and critical phenomena relies on the availability of well-defined model systems. In particular, a targeted synthesis and characterization of materials with interaction parameters that closely approximate those of theoretical model systems and yield energy scales of temperature and magnetic field that are accessible by existing experimental infrastructures is required. The effective Hamiltonian to describe a quasi-2D QHAF in an applied magnetic field is

$$\mathcal{H} = J \sum_{i,j} \left[ S_i^x S_j^x + S_i^y S_j^y + (1-\Delta) S_i^z S_j^z \right] \\ + J' \sum_{i,i'} \mathbf{S}_i \mathbf{S}_{i'} - g\mu_B \mu_0 H \sum_i \mathbf{S}_i, \quad (1)$$

where $J$ and $J'$ are the intra- and interlayer exchange couplings, and $\Delta$ scales the deviation from an ideal Heisenberg interaction to an easy-axis or easy-plane characteristic. The first sum in Eq. (1) is taken over nearest-neighbor (NN) spins in the quasi-2D planes, the second summation is over NN spins in adjacent layers, and the Zeeman term applies to all moments. For external magnetic fields smaller than the anisotropy field $H_A$ and $T \ll J/k_B$, the formation of anisotropic magnetic correlations is driven by the intrinsic anisotropy $\Delta$. Conversely, for fields exceeding $H_A = \Delta \times H_{sat}$, where $H_{sat}$ denotes the saturation field, the anisotropy of magnetic correlations is mainly determined by the strength and direction of the applied magnetic field. In the Hamiltonian (1), a positive $J$ corresponds to an antiferromagnetic exchange, and $0 < \Delta \le 1$ describes the degree of easy-plane, or $XY$, anisotropy.

* Corresponding author. E-mail: d.dmytriieva@hzdr.de
† Corresponding author. E-mail: h.kuehne@hzdr.de
‡ Corresponding author. E-mail: CLandee@clarku.edu

The discovery of the high-temperature cuprate superconductors attracted great attention to the 2D spin-1/2 Heisenberg model [8–16]. Some of the undoped parent compounds, such as $Sr_2CuCl_2O_2$ and $La_2CuO_4$, are known to be excellent realizations of 2D spin-1/2 Heisenberg antiferromagnets with large nearest-neighbor exchange couplings of the order of 1000 K [17–20]. However, in contrast to the ideal 2D Heisenberg model, a transition to a Néel-type state at finite temperatures is observed for all materials reported up to now. This transition is often discussed in terms of finite interlayer interactions [21–24], spin anisotropy [3, 25, 26], dipolar anisotropy, and other symmetry-allowed contributions to the Hamiltonian [27].

Whereas the critical phenomena of several systems were mapped onto the BKT theory, such as the superfluid transition of thin $^4$He films [28], the solid-on-solid model [29], the two-dimensional melting [30], the superconducting transition of Josephson junctions [31], the collision physics of 2D atomic hydrogen [32], the loss of quasi-coherence of a trapped degenerate quantum gas of rubidium atoms [33], and the magnetic van der Waals antiferromagnet in the atomically thin limit [34], a very clean bulk realization of the square 2D $XY$ Heisenberg lattice is lacking up to now. There are two main obstacles for the experimental realization of a 2D $XY$ antiferromagnet. One challenge relates to the unavoidable existence of finite interlayer interactions. In case of the 2D Heisenberg magnets, even an infinitesimal interlayer interaction is sufficient to perturb the critical behavior of the system and stabilize a conventional type of Néel order below a transition temperature $T_N$ [35]. Secondly, a weak intrinsic easy-plane anisotropy constraints the temperature range of $XY$-type correlations. For the case of crystalline magnetic lattices of $Cu^{2+}$ ions, the exchange coupling between spin moments as well as the single-ion properties are almost isotropic.

Another approach to realize a well-defined investigation of the magnetic correlations in low-dimensional spin systems is by tuning the Zeeman terms of the effective Hamiltonian with the application of an external magnetic field. This gives rise to the unique possibility of probing a quasi-2D spin system with well-defined $XY$ anisotropy in the experiment. It was shown by quantum Monte Carlo calculations that the application of a magnetic field to an isotropic 2D Heisenberg antiferromagnet can be mapped onto the anisotropic 2D $XY$ model in zero field, where the strength of the spin exchange anisotropy can be continuously tuned by the external field [36, 37]. This overall context has triggered extensive and ongoing efforts to synthesize novel quasi-2D Heisenberg model materials with highly isolated layers and relatively small antiferromagnetic interaction energies, allowing for the investigation of field-induced effects in moderate applied magnetic fields. Molecular-based materials consisting of $3d$ transition metal ions, such as copper, embedded into an organic matrix, are of particular interest. The combination of different ligands gives the opportunity to engineer a wide range of materials with well-defined magnetic properties [38–48].

In this paper, we report a comprehensive investigation of the magnetic properties of the newly synthesized compound $[Cu(pz)_2(2\text{-HOpy})_2](PF_6)_2$ (CuPOF), where $pz = C_4H_4N_2$ and 2-HOpy $= C_5H_4NHO$. The material is characterized by magnetometry, ESR, specific heat, $\mu^+SR$, and NMR. CuPOF is shown to be a very clean realization of the square 2D spin-1/2 Heisenberg lattice with moderate intralayer antiferromagnetic NN exchange, $J/k_B = 6.80(5)$ K, and highly-isolated magnetic layers, with $J'/J \simeq 10^{-4}$. A weak intrinsic easy-plane anisotropy, revealed by bulk magnetometry, yields a temperature-driven crossover of the spin correlation anisotropy from isotropic Heisenberg to anisotropic $XY$-type behavior, which, under the influence of a finite interlayer coupling $J'$, constitutes a driving mechanism for a transition to long-range commensurate order. A strong increase of the transition temperature upon application of magnetic field from 1.38(2) K at zero field to 2.8 K at 7 T is caused by the field-driven increase of the anisotropy of spin correlations.

## II. EXPERIMENTAL

The compound $[Cu(pz)_2(2\text{-HOpy})_2](PF_6)_2$ was synthesized using conventional solution chemistry techniques, described in detail in the Supplemental Material (SM) [49]. Slow evaporation of methanol solutions of the product, [Cu(pz)$_2$(2-HOpy)$_2$](PF$_6$)$_2$, (pz = pyrazine, 2-HOpy = 2-pyridone) produces thin, rectangular, green plates, typically 3 to 5 mm on a side and about 1 mm thick. The crystals extinguish well under polarized light and are dichroic.

X-ray data were obtained on an Agilent Technologies Gemini Eos CCX-ray diffractometer using Cu-$K\alpha$ radiation ($\lambda$ = 1.5418 Å) with $\omega$ scans using CrysAlisPro software [50] refined cell parameters and SCALE3 AB-SPACK [51] scaling algorithm defined absorption corrections. Data were collected at 120 K. SHELXS97 [52] was used to solve the structures, which were refined via least-squares analysis using SHELXL-2016 [53]. All non-hydrogen atoms were refined anisotropically. Hydrogen atoms bonded to nitrogen atoms were located in the difference Fourier maps and their positions refined with fixed isotropic thermal parameters. The remaining hydrogen atoms were geometrically located and refined using a riding model with fixed isotropic thermal parameters. The structure has been deposited with the Cambridge Crystallographic Data Centre (CCDC) (1553982). A Bruker D8 powder x-ray diffractometer was used to confirm the purity and phase of powdered samples prior to magnetic measurements.

Measurements of magnetic bulk properties between 1.8 and 310 K were carried out using a Quantum Design MPMS-XL SQUID magnetometer with a 5 T magnet, as well as a vibrating sample magnetometer (VSM) Quantum Design PPMS with a 14 T magnet. Corrections were made to the data for the background signal of the sample holder, as well as diamagnetic contributions. Studies below 2 K were performed employing a $^3$He cooling stage.

The high-field magnetization of CuPOF single crystals in pulsed fields up to 35 T and at temperatures of 0.37 and 1.4 K were performed at the Dresden High Magnetic Field Laboratory (HLD) at the Helmholtz-Zentrum Dresden-Rossendorf. Additional measurements of a polycrystalline sample in DC fields up to 35 T were done using a vibrating sample magnetometer at the National High Magnetic Field Laboratory (NHMFL) in Tallahassee, as well as in pulsed fields up to 25 T at the NHMFL facility in Los Alamos. The results are fully consistent with the data from the HLD and can be found in the SM.

Room-temperature electron spin resonance (ESR) studies were performed on polycrystalline material and single crystals of CuPOF, using a commercially available X-Band Bruker ESR spectrometer operating at 9.8 GHz at Clark University. EasySpin [54] was used to determine the $g$-factors and linewidths. Additional ESR measurements of angular-dependent spectra, as well as temperature-dependent spectra for field parallel to the $c$ axis, were performed at the HLD between 3 and 300 K at 9.4 GHz, employing an X-band Bruker ELEYSYS E500 ESR spectrometer. The obtained values of the anisotropic $g$-factor are consistent with the results measured at Clark University, and are presented in the SM. High-frequency ESR measurements along the crystallographic $c$ axis at 1.5 K and fields up to 16 T were per-formed at the HLD using a home-built transmission-type tunable-frequency ESR spectrometer, similar to that described in Ref. [55], with a probe in Faraday configuration. These results can as well be found in the SM.

Heat-capacity measurements between 1.8 and 300 K were performed using a Quantum Design PPMS system. Further, a $^3$He insert was used to record the heat capacity at temperatures down to 0.4 K. Powdered samples with masses of 1.065(5) mg and 1.832(5) mg, for measurements at $^4$He and $^3$He temperatures, respectively, were pressed into pellets and attached to the sample platforms using Apiezon N grease. The addenda was determined from measurements with an empty sample holder, and subtracted from the data to obtain the heat capacity of the sample.

Zero-field muon-spin relaxation ($\mu^+$SR) measurements on a polycrystalline sample were carried out using the EMU spectrometer at the ISIS facility at the Rutherford Appleton Laboratory. The sample was mounted on an Ag backing plate and covered with a 12.5 $\mu$m thick Ag foil mask before being inserted into a $^3$He/$^4$He cryostat. Further technical details are provided in the SM.

$^1$H nuclear magnetic resonance (NMR) spectra were recorded using a commercial solid-state spectrometer. A standard Hahn spin-echo pulse sequence with stepped sweep of the carrier frequency was employed to record the broad-bandwidth spectra. The NMR probe was equipped with a single-axis goniometer for precise orientation of the magnetic field parallel to the crystallographic $c$ axis. The measurements at 1.6 K and above were performed in a 8 T high-resolution magnet equipped with a $^4$He flow cryostat.

## III. RESULTS

*Crystal Structure.* [Cu(pz)$_2$(2-HOpy)$_2$](PF$_6$)$_2$ crystallizes in the orthorhombic space group *Cmca*. The asymmetric crystallographic unit comprises one Cu$^{2+}$ ion, two half-pyrazine molecules, one 2-pyridone molecule and two PF$_6^-$ ions. The local coordination sphere of the Cu$^{2+}$ ion is presented in Fig. S1 in the SM. The Cu$^{2+}$ ion sits on a two-fold axis (parallel to $b$), one coordinated, dissymmetric (N11/N14) pyrazine sits on the same two-fold axis while the other pyrazine molecule (N21) lies across a mirror plane normal to the $a$ axis. The two PF$_6^-$ ions also lie on mirror planes such that there are five independent fluoride ions in each. The Cu$^{2+}$ coordination sphere exhibits a classic distorted Jahn-Teller octahedral geometry with four bridging pyrazine molecules in the equatorial plane [Cu–N = 2.05(1) Å] and elongated Cu–O bonds [2.285(1) Å] in the axial sites. The copper ion and all four bound nitrogens are co-planar as required by symmetry. The Cu–O1 bond is nearly perpendicular to the plane, making an angle of 1.0° with the normal to the plane. Crystal data and structure refinement parameters, as well as selected bond lengths and angles of CuPOF, are presented in Tables S1 and S2 in the SM.

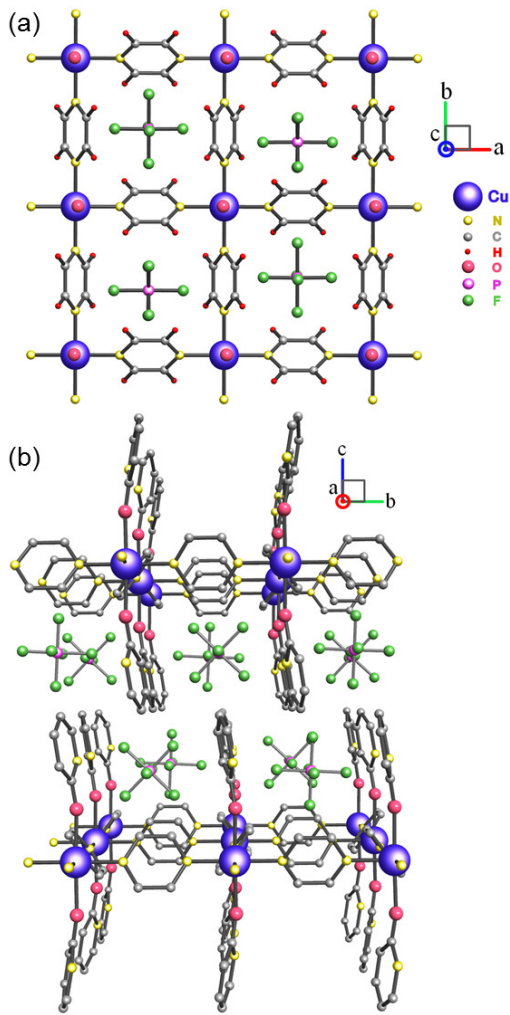

(a)

Cu
O
N
C
H
O
P
F

(b)

Figure 1. (a) View of the Cu-pyrazine layers of CuPOF along the $c$ axis. The 2-pyridone molecules extend normal to the planes and are coordinated to the copper atoms through the oxygens. The uncoordinated $PF_6^-$ anions occupy vacancies in the lattice. (b) View of CuPOF along the $a$ axis. The Cu-pyrazine layers extend horizontally and into the page. The interdigitation of the 2-pyridone molecules leads to the extreme isolation of the layers. The 2-pyridone molecules within one layer are all canted in the $+b$ direction while the molecules in the adjacent layers cant in the $-b$ direction.

The bridging pyrazine ligands link the $Cu^{2+}$ ions into a nearly square layer, see Fig. 1(a). The pyrazine rings exhibit a propeller twist relative to the Cu plane, with the N11 rings canted $66.9°$ and the N21 rings canted $53.1°$ out of the plane. This results in slightly different Cu–Cu distances of $6.676$ Å through the N11 ring and $6.680$ Å through the N21 ring. The layers are further separated by the $PF_6^-$ anions, which are located in the pockets between the 2-pyridone ligands, resulting in a minimum Cu–Cu distance in adjacent layers of $13.097$ Å.

The structure of CuPOF possesses a hidden canting, see Fig. 1(b). The Cu-pyrazine layers lie in the $ab$ plane (into the page and horizontal) while the Cu-oxygen bonds are nearly parallel to the $c$ axis but canted by $±1.0°$ towards $b$. Within each layer, the canting is in the same direction but adjacent layers are canted in the opposite direction. A similar canting is seen in the orientation of the pyridone rings; they are alternately tilted by $8.7°$ away from the normal to the Cu-pyrazine planes. The chemical equivalence of the 2-pyridone molecules is confirmed by our $^{13}C$ magic-angle spinning (MAS) NMR spectroscopy (see Fig. S11 in the SM).

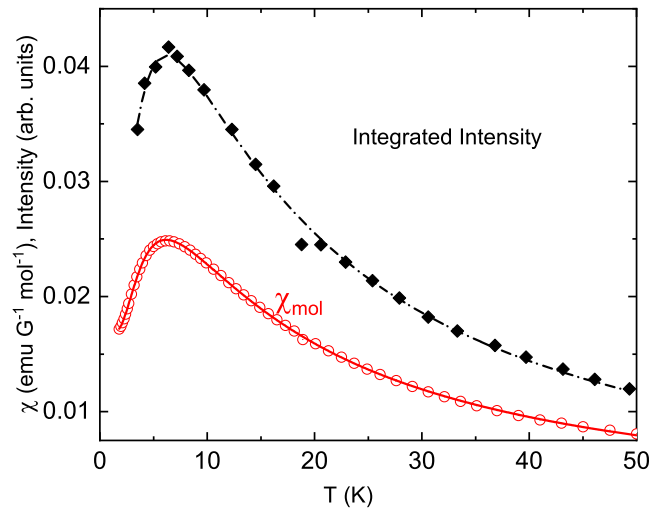

Figure 2. Temperature dependence of the powder susceptibility (circles) in $emu G^{-1} mol^{-1}$ and the integrated ESR intensity at 9.4 GHz (diamonds) in arbitrary units. The red line shows the best fit of the 2D QHAF model calculations to the magnetic susceptibility of a polycrystalline sample. The black dash-dotted line represents the fit of the same model to the integrated ESR intensity of a polycrystalline CuPOF sample.

*Magnetometry.* The magnetic susceptibility of a polycrystalline sample of CuPOF is shown in Fig. 2, and yields a rounded maximum near $6.8$ K. At higher temperatures ($T > 40$ K), the data are well described by a Curie-Weiss law with a Curie constant of $0.440(5)$ $emu G^{-1} mol^{-1} K$ and a Curie-Weiss temperature $\Theta_{CW} = -5.2(6)$ K, indicating a small antiferromagnetic interaction. Accordingly, the data were compared to the susceptibility of a 2D $S = 1/2$ Heisenberg antiferromagnet [56] in which the Curie constant, exchange strength $J$, and small paramagnetic impurity fraction were adjusted. The best agreement between data and the model calculations, denoted by the red curve in Fig. 2, is obtained with $C = 0.445(5)$ $emu G^{-1} mol^{-1} K$, $J/k_B = 6.80(5)$ K, and $1.1(1)\%$ paramagnetic contribution. The agreement between the Curie constants from the Curie-Weiss model and the 2D QHAF model is excellent, with their values corresponding to an average $g$-value of $2.17$. The average $g$-factor has been determined at room temperature using ESR, and found to be $\langle g \rangle = 2.15$.

The existence of two distinct pyrazine molecules in the unit cell allows for the possibility of a rectangular magnetic lattice in which the exchange strengths ($J$ and $\alpha J$, $0 \leq \alpha \leq 1$) along the $a$ and $b$ axes are different. This possibility has been tested by comparing the susceptibility data to the susceptibilities of rectangular 2D QHAF [57, 58] for various values of $J$ and $\alpha$. The square lattice ($\alpha = 1$) case gives by far the best fit and it is possible to rule out any rectangular contribution with $\alpha < 0.96$.

The low-temperature static susceptibility of a single crystal of CuPOF at 0.1 T is shown in the inset of Fig. 3. The out-of-plane DC susceptibility ($H \parallel c$) has a minimum at 1.86(5) K, whereas for a field applied in the $ab$ plane ($H \perp c$), the DC susceptibility steadily decreases to the lowest measured temperature. The minimum in the out-of-plane susceptibility at $T_{co}$ indicates the presence of an $XY$ anisotropy, where, with decreasing temperature, the correlation of spin moments crosses from isotropic to easy-plane behavior. The temperature-dependent out-of-plane susceptibility at different magnetic fields is presented in the main panel of Fig. 3. With increase of the magnetic field, the broad minimum of the static susceptibility shifts to higher temperatures, as indicated by the triangles. At fields above around 4 T, the minimum broadens, and cannot be observed anymore.

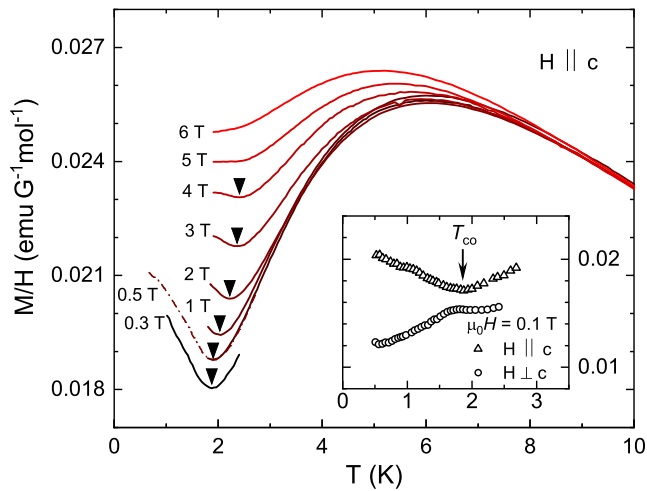

Figure 3. Temperature dependence of the out-of-plane molar susceptibility of a single-crystalline sample of CuPOF at different magnetic fields. The solid black triangles in the main panel indicate the crossover temperature $T_{co}$, as discussed in the main text. Inset: The low-temperature susceptibility of a single crystal of CuPOF. The data were collected in a field of 0.1 T. The susceptibilities normal to the plane and within the $ab$ plane are represented by the open triangles and circles, respectively. The vertical arrow indicates the crossover temperature $T_{co}$ for $M/H \parallel c$ at 1.86(5) K. The broad anomaly of the in-plane static susceptibility at about 1.6(1) K is attributed to a background contribution.

The magnetization of a single crystal of CuPOF has been measured up to 1 T at 0.5 K, see insets of Figs. 4(a) and 4(b). For a field parallel to the $c$ axis, the magnetization monotonously increases over that range. In contrast, when the field is applied in the $ab$ plane ($H \perp c$), a small spin-flop anomaly is observed at around 0.36(1) T.

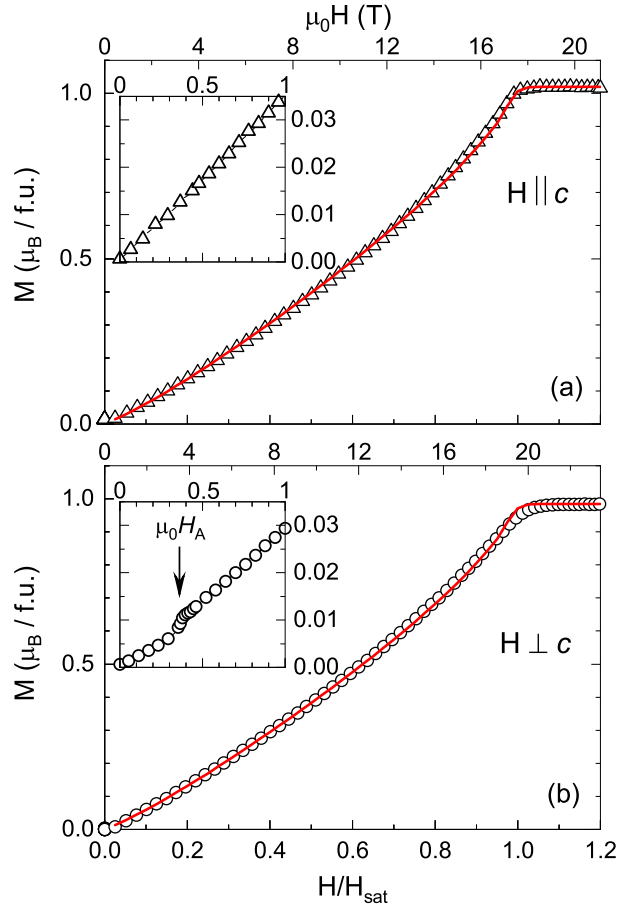

Figure 4. Magnetization of single-crystalline CuPOF at 0.37 K in (a) out-of-plane and (b) in-plane direction and comparison to the respective QMC calculations (red line). The insets show the magnetization at 0.5 K in magnetic fields up to 1 T in enlarged scale. The vertical arrow indicates the anisotropy field at $\mu_0 H_A = 0.36(1)$ T. The bottom axis represents the relative magnetic field $H/H_{sat}$. The corresponding absolute values of the magnetic fields are shown in the upper axis.

The relative magnetization of single crystals of CuPOF was determined at several temperatures for fields up to 35 T both parallel to the $c$ axis ($H \parallel c$), and within the layers ($H \perp c$). The experimental values at 0.37 K are presented in panels (a) and (b) of Fig. 4. The absolute values of the magnetization were obtained from a direct comparison with the magnetization recorded at 4 K for $H \parallel c$ (at 2 K for $H \perp c$) up to 14 T, using a VSM magnetometer, see Fig. S3 in the SM. For both magnetic-field orientations, the saturated moment is about 1 $\mu_B$ per formula unit, as expected for the $Cu^{2+}$ ion. The experimental data (symbols) are compared to quantum Monte Carlo (QMC) simulations (lines) for the 2D QHAF at the relative temperature $k_B T / J = 0.05$. Excellent agreement between the experimental data and the QMC simulations is found for both field orientations. Over the full field range, the deviation between experimental and theoretical data is below ±1% for $H \parallel c$ and below ±2% for $H \perp c$. The in- and out-of-plane saturation fields were determined as 17.6 and 19.6 T for $H \parallel c$ and $H \perp c$, respectively. In mean-field approximation, the saturation field $H_{\text{sat}}$ is defined by the exchange strength $J$,

$$g\mu_B\mu_0 H_{\text{sat}} = 2zJS, \qquad (2)$$

where $z = 4$ is the number of nearest-neighbor magnetic moments. The exchange strength can be computed from Eq. (2) using the experimentally determined saturation fields and the respective $g$-values, $g_c = 2.29(1)$ at 1.5 K and $g_{ab} = 2.070(7)$ at room temperature, see ESR section below. The obtained values, $J/k_B = 6.75(5)$ and 6.78(5) K, are in very good agreement with each other and those determined from the susceptibility and specific-heat measurements.

*Electron Spin Resonance.* The anisotropy of the room-temperature ESR spectrum of a single-crystalline CuPOF sample was investigated at 9.8 GHz. A single ESR line was observed for each field orientation, see Fig. S6 in the SM. The resonance field, used to extract the electronic $g$-factor and the spectral linewidth, was determined by modeling a nearly Lorentzian function to the experimentally obtained spectra [54, 59, 60]. The angular dependence of the $g$-factor in the $ac$, $bc$, and $ab$ planes indicates a strong planar-like anisotropy, see Fig. S7 in the SM, with $g_a = 2.073(2)$, $g_b = 2.066(4)$, and $g_c = 2.298(2)$ [61].

The temperature-dependent integrated ESR-line intensity, scaling as the bulk susceptibility of the sample [62] is shown in Fig. 2 for temperatures between 3 and 50 K with field applied along the $c$ axis. Analogously to the modeling of the DC susceptibility, the integrated ESR intensity was modeled, varying only the Curie constant, with fixed exchange coupling of $J/k_B = 6.80(5)$ K and a paramagnetic impurity percentage of 1.1(1)% (black dash-dotted line in Fig. 2).

High-frequency ESR spectroscopy at 1.5 K in the frequency range between 52 and 500 GHz revealed a single resonance mode with a linear frequency-field dependence (Fig. S9 in the SM), yielding $g_c = 2.29(1)$.

*Specific Heat.* The specific heat was measured between 0.4 and 300 K. The data increase smoothly from about 0.2 Jmol⁻¹K⁻¹ at 1 K, approaching 640 Jmol⁻¹K⁻¹ at 300 K. No sharp anomalies, corresponding to structural changes or ordering transitions, were observed in this range (Fig. S10 in the SM). The data between 1 and 9 K are shown in Fig. 5(a) (black open circles) revealing a broad hump exceeding the normal phonon contribution. The data in this temperature range were analyzed as a sum of the magnetic specific heat of a 2D QHAF and a phononic contribution. The low-temperature lattice contribution to the specific heat, stemming from a complex phononic spectrum in the molecular-based material CuPOF, is best approximated by choosing a three-term polynomial with $C_{\text{pho}} = AT^3 + BT^5 + CT^7$. The magnetic specific heat was represented as a ratio of polynomials, similar to the approach used in a previous study [63], but is based on recent QMC simulations of the magnetic specific heat [64] that extended to lower relative temperatures. The range of validity for the square lattice is between $0.15 \leq k_B T/J \leq 5.0$ (see SM and Table S4 therein).

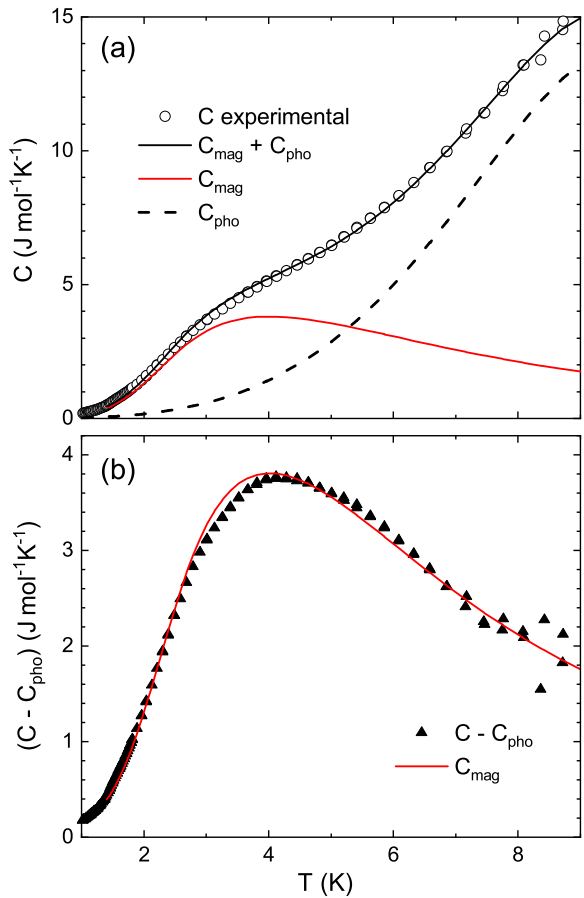

FIG. 5. (a) Temperature dependence of the normalized specific heat of CuPOF denoted as open black circles, with the black line representing the best fit to these data by a sum of magnetic (solid red line) and phonon (black dashed line) contributions, $C_{\text{pho}}$. (b) The black solid triangles represent $C_{\text{mag}}$, the difference between $C$ and $C_{\text{pho}}$.

The resulting best fits to $C_{\text{pho}}$ and $C_{\text{mag}}$ are shown in Fig. 5(a). The sum of the two individual contributions appears as the solid black line and shows very good agreement with the experimental data. The modeling parameters are $J/k_B = 6.75(5)$ K, $A = 20.2 \times 10^{-3}$ Jmol⁻¹K⁻⁴, $B = 16.6 \times 10^{-5}$ Jmol⁻¹K⁻⁶, and $C = -23.7 \times 10^{-7}$ Jmol⁻¹K⁻⁸. As seen in Fig. 5(a), the magnetic contribution dominates at lower temperatures. Including data at higher temperatures in the fitting process did not change the value of the exchange strength, since $C_{\text{mag}}$ is a rapidly decreasing fraction of the total specific heat. Fig. 5(b) shows a direct comparison of the calculated magnetic specific heat (red line) to the difference between the total and estimated lattice specific heat (black triangles). The obtained exchange strength of $J/k_B = 6.75(5)$ K is in excellent agreement with that obtained from the susceptibility and magnetization results. Similar to the analysis of the bulk susceptibility, the specific-heat data were investigated in terms of the possibility of a rectangular magnetic lattice [64]. The results are consistent with a magnetic square lattice.

*Muon Spin Relaxation.* Zero-field muon-spin relaxation ($\mu^+$SR) spectra [65] for CuPOF are shown in Fig. 6(a). Spontaneous oscillations in the asymmetry function $A(t)$, which is proportional to the spin polarization of the muon ensemble, see SM, were observed at low temperatures. An oscillating behavior of the muon spin polarization is characteristic of the presence of quasistatic long-range magnetic order (LRO). The local magnetic field that results from LRO causes those muons with spin perpendicular to the local field to precess coherently at the frequency $\nu_i$, where $\nu_i$ is proportional to the magnitude of the local field $B_i$ at the $i$th muon site. Changes in the spectra are observed in measurements across the temperature range $0.1 < T < 2$ K, which is parametrized by modeling the asymmetry $A(t)$ with the relaxation function

$$A(t) = (A_0 - A_{bg})\cos(2\pi\nu t)\,e^{-\lambda_1 t} + A_{bg}e^{-\lambda_2 t}, \quad (3)$$

where $A_0$ is the initial asymmetry at $t = 0$, the parameters $\nu$ and $\lambda_1$ are the precession frequency and relaxation rate of the oscillatory component, respectively. The parameters $A_{bg}$ and $\lambda_2$ account for the background muons and those with spin initially parallel to the local magnetic field.

The evolution of the parameters $A_0$, $\lambda_1$, and $\nu$ are shown in Figs. 6(b)-6(d). The precession frequency, see Fig. 6(d), which is proportional to the magnitude of internal magnetic field probed by the muon spins, shows a monotonic decrease from base temperature up to 1.4 K, but starts to rise again before becoming roughly constant above 1.5 K. A sharp maximum of $A_0$ and discontinuity of $\lambda_1$ are observed just below $T = 1.4$ K, coincident with the minimum of $\nu$.

The behavior of $\nu$, together with the peak of $A_0$ and the discontinuity of $\lambda_1$, suggest that CuPOF undergoes a magnetic phase transition around 1.4 K. However, unlike many magnetic systems where the precession vanishes above $T_N$, the asymmetry still appears to show oscillatory behavior. Such an oscillatory signal is common in similar materials containing fluorine nuclei in the paramagnetic phase [66]. It arises because the electronic moments fluctuate outside of the muon time window and are consequently removed from the spectrum, leaving the muon sensitive to the nuclear magnetic moments. In fluorinated materials, there are frequently muon sites where the positive muon sits close to the electronegative fluorine and enters a dipole-dipole coupled entangled state, leading to heavily damped, low-frequency oscillations [66]. A similar scenario for CuPOF is suggested, allowing identification of the antiferromagnetic transition temperature with the discontinuities in the modeled parameters.

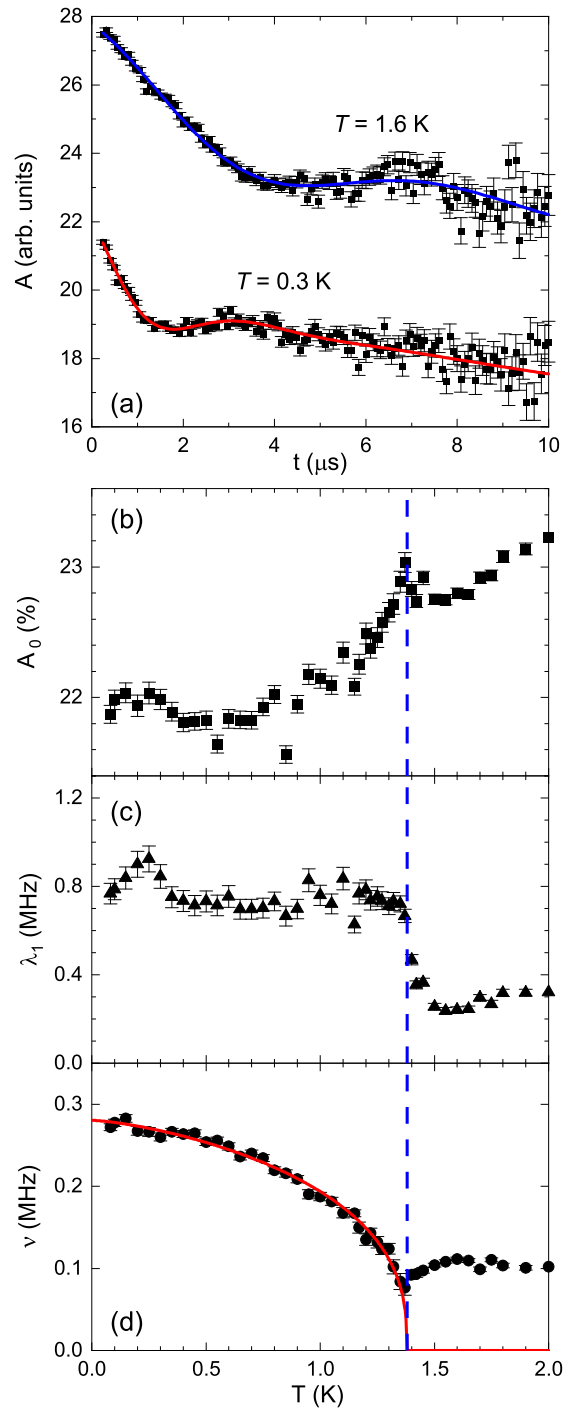

Figure 6. (a) Representative $\mu^+$SR spectra of CuPOF at 0.30 and 1.60 K. Solid lines are fits to the data using Eq. 3. (b-d) Temperature dependence of the parameters $A_0$, $\lambda_1$ and $\nu$. The blue dashed line indicates $T_N = 1.38(2)$ K.

To study the critical behavior close to the phase transition, we exclude the data above 1.4 K. To extract the critical exponent of the order parameter, the temperature dependence of the precession frequency was modeled by

$$\nu(T) = \nu(0) \left[ 1 - \left( \frac{T}{T_N} \right)^\alpha \right]^\beta. \tag{4}$$

The best fit with this model gives an ordering temperature $T_N = 1.38(2)$ K, a critical exponent $\beta = 0.37(2)$, and a phenomenological parameter $\alpha = 1.4(2)$. The resulting curve is shown in Fig. 6(d). Fixing $\alpha$ to unity and fitting the data between 0.85 and 1.40 K, i.e. closely below the transition temperature, results in very similar values for the critical exponent $\beta = 0.344(30)$ and the transition temperature $T_N = 1.382(10)$ K. Note that the critical temperature is consistent with the discontinuities found in Figs. 6(b) and 6(c).

*Nuclear Magnetic Resonance.* Selected $^1$H-NMR spectra, recorded for an out-of-plane field of 7 T and temperatures in the regime of long-range order, are presented in Fig. 7(a). Due to several non-equivalent hydrogen sites in the crystallographic unit cell, the resulting $^1$H-NMR spectrum is composed of many resonance peaks, and, therefore, rather complicated. Since the same qualitative temperature dependence was observed for all $^1$H lines of the spectrum, we consider, in the following, only selected lines with comparably little overlap. At temperatures above $T_N$ and an out-of-plane field of 2 T, a single, slightly non-symmetric Gaussian-like line is observed. The larger field of 7 T allows one to assign two non-equivalent hydrogen sites with Gaussian line shape in the paramagnetic regime, as exemplified by the blue and red fits to the spectrum at 3.5 K in the inset of Fig. 7(c).

As shown in Fig. 7(b) and 7(c), with decreasing temperature, the $^1$H-NMR line splits into two sets of doublets, revealing a phase transition to long-range order at 2.3 and 2.8 K for $\mu_0 H \parallel c = 2$ and 7 T, respectively. The splitting of the NMR spectrum is a clear signature of commensurate antiferromagnetic order, where each of the two lines represents a sublattice magnetization of opposite local spin polarization. The observation of multiple doublets is caused by several non-equivalent hydrogen sites in the lattice, with coincidental overlap of the nuclear resonance frequency in the paramagnetic temperature regime. Due to the different hyperfine coupling constants of the corresponding $^1$H sites, this overlap is lifted in the ordered state. Considering the quasi-2D, almost-square structure of the $Cu^{2+}$ ions in CuPOF, the commensurate antiferromagnetic order is presumably of checkerboard type.

We note that the temperature dependence of the sublattice magnetization curves deviates from the mean-field type behavior probed by the $\mu^+$SR precession frequency at zero field. The details of this field-induced behavior will be a subject of future, more detailed investigations by local-probe techniques.

As part of a thorough characterization of CuPOF, additional room-temperature $^{13}$C magic-angle spinning (MAS) NMR and $^{31}$P cryo-MAS NMR studies [67] were performed, and are presented in SM (Figs. S11-S15).

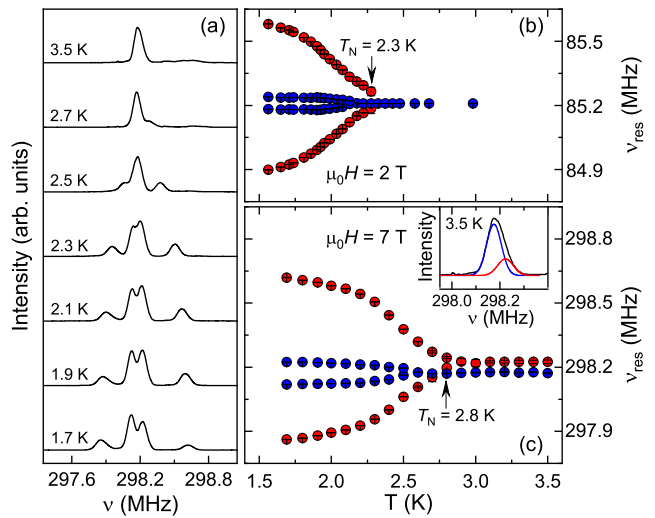

Figure 7. (a) $^1$H-NMR spectra of CuPOF at selected temperatures in the regime of magnetic order, recorded at $\mu_0 H \parallel c = 7$ T. Temperature-dependent peak positions $\nu_{res}$ at out-of-plane fields of (b) 2 and (c) 7 T. Vertical arrows mark the onset temperature of LRO. The inset of (c) shows a typical $^1$H-NMR spectrum at 7 T and 3.5 K. Two non-equivalent hydrogen sites can be assigned in the paramagnetic regime.

## IV. DISCUSSION

The ideal 2D QHAF is described by the Hamiltonian Eq. (1) for the case $J' = \Delta = 0$. Without applied field, the thermal fluctuations at arbitrarily low temperatures prevent a semi-classical order, regardless of the strength of the intralayer interaction $J$ [4]. Any small perturbations, such as a finite interlayer coupling or spin-exchange anisotropy, give rise to quasi-long-range order below a non-zero transition temperature. Hence, the ratio $k_B T_N / J$ can be treated as a measure of perturbations to the pure 2D QHAF [35]. It varies between zero for the 2D and 0.946(1) for the 3D isotropic spin-1/2 Heisenberg model [68]. For typical molecular-based Cu-pyrazine materials, the reported values of $k_B T_N / J$ are between about 0.2 and 0.3, see Table I. The lowest ratio $k_B T_N / J = 0.176$ yet found applies to the inorganic material $Sr_2CuO_2Cl_2$, with a strong intralayer coupling of $J / k_B \simeq 1450$ K.

The nature of the transitions induced by finite $J'$ or $\Delta$ are fundamentally different. Whereas the interlayer coupling $J'$ induces a Néel state [35], a finite easy-plane or $XY$-like anisotropy yields a temperature-driven crossover of the spin correlations from isotropic Heisenberg to anisotropic $XY$-type behavior at a finite temperature $T_{co}$, predicted to lead to a topological BKT transition at $T_{BKT} < T_{co}$ [25, 26, 69].

In all bulk materials, finite interlayer couplings and magnetic anisotropies are present, see Table I for some selected cases. However, until now, there has been no theoretical framework that allows for an accurate experimental determination of $J'$ in the presence of a non-zero anisotropy $\Delta$ [40]. Still, the challenges associated with a clean experimental realization of the 2D QHAF can be discussed as follows. Predicted critical temperatures for the two limiting cases of a finite $J'$ with $\Delta = 0$, as well as that of a finite $\Delta$ with $J' = 0$ are presented in Fig. 8 as functions of the perturbation parameter, $J'/J$ or $\Delta$, respectively. The black curve represents the normalized Néel temperature, $k_\mathrm{B}T_\mathrm{N}/J$, of a 3D array of isotropic square 2D spin-1/2 Heisenberg planes with a coupling $J'$ between adjacent layers as a function of the exchange ratio $J'/J$ from Ref. [35].

$$k_\mathrm{B}T_\mathrm{N} = \frac{4\pi\rho_s}{2.43 - \ln(J'/J)}, \qquad (5)$$

where $\rho_s = 0.183J$ is the renormalized spin-stiffness constant. Note that the vertical axis is linear, whereas the horizontal axis is logarithmic and spans three orders of magnitude. The very weak decrease of the ordering temperature with reduction of the ratio $J'/J$ results from the exponential divergence of the correlation length of the 2D QHAF at low temperatures [17, 70].

In the other limiting case with $J' = 0$ and finite $\Delta > 0$, quantum Monte Carlo calculations showed that even for anisotropies as small as $10^{-3}$, the critical behavior of the magnetic lattice resembles that of the Berezinskii-Kosterlitz-Thouless universality class. A slow logarithmic decrease of the $T_\mathrm{BKT}$ temperature with reduction of the spin-exchange anisotropy was determined as [69]

$$k_\mathrm{B}T_\mathrm{BKT} = \frac{2.22J}{\ln(330/\Delta)}, \qquad (6)$$

as depicted by the red dashed line in Fig. 8.

From the comparison of these results, it is apparent that the interlayer interaction determines the ordering process for equal magnitudes of $J'/J$ and $\Delta$. Upon cooling from $T > J/k_\mathrm{B}$, the onset of the 3D long-range order occurs before any signatures of the exchange anisotropy can be observed. The influence of the spin anisotropy is only relevant if the interlayer interaction $J'/J$ is significantly smaller than $\Delta$. Therefore, minimizing the interlayer spin interaction is of key importance for developing any material approximation to the ideal 2D QHAF.

In contrast to a strongly anisotropic spin system with $\Delta \simeq 1$, where the topological excitations of unpaired vortices and antivortices are formed well above $T_\mathrm{BKT}$, a qualitatively different behavior is expected for a weakly anisotropic spin system with $\Delta \ll 1$ [26]. At high temperatures, the spin correlations can well be approximated as isotropic. With decreasing temperatures, the $XY$ anisotropy becomes relevant and stabilizes a planar spin configuration. The formation of vortices and antivortices starts in the regime of the temperature $T_\mathrm{co}$, which indicates the crossover between isotropic and $XY$ behavior. Quantum Monte Carlo calculations revealed that the uniform susceptibility is very sensitive to this crossover phenomenon [26, 69, 71]. Whereas the in-plane susceptibility monotonously decreases with temperature below about $J/k_\mathrm{B}$, a characteristic minimum of the out-of-plane susceptibility marks the crossover from isotropic to anisotropic behavior at $T_\mathrm{co}$. The dependence of the crossover temperature on the spin anisotropy $\Delta$ can be described by the empirical expression [26]

$$k_\mathrm{B}T_\mathrm{co} = \frac{2.69J}{\ln(160/\Delta)}, \qquad (7)$$

and is depicted by the dash-dotted blue line in Fig. 8.

Furthermore, in the presence of $XY$ spin anisotropy, the field-dependent magnetization is expected to yield a qualitatively different behavior for in- and out-of-plane orientations of the field at $T < T_\mathrm{co}$. Below $T_\mathrm{co}$, the antiferromagnetically coupled magnetic moments preferably fluctuate in the easy plane. The application of a magnetic field suppresses longitudinal spin fluctuations, effectively inducing an easy-plane anisotropy of spin correlations perpendicular to the field direction. For a field applied perpendicular to the intrinsic easy plane, this yields a monotonous increase of the magnetization. On the other hand, for a magnetic field applied in plane, when the Zeeman term becomes larger than the intrinsic anisotropy energy, the total spin polarization in the field direction will enhance, yielding a slope increase of the magnetization curve at the anisotropy field $H_\mathrm{A}$. Therefore, $H_\mathrm{A}$ represents a measure of the spin-anisotropy parameter, and can be estimated as $\Delta = H_\mathrm{A}/H_\mathrm{sat}$.

It is worth noting, that for most bulk realizations of the 2D QHAF model, the weak intrinsic $XY$ anisotropies do not stem from anisotropic exchange interactions. For the prevalent case of 2D QHAF materials based on $Cu^{2+}$ ions, the exchange between spins is almost isotropic. Typically, a weak anisotropy of a few percent of the isotropic superexchange interactions stems from crystal electric field effects and related non-quenched orbital contributions. Commonly, the magnetic properties are then described by an effective-spin formalism, yielding an anisotropic $g$-factor. Equivalently, this anisotropy can be treated in terms of an anisotropic exchange of isotropic effective spin moments, as described by the Hamiltonian Eq. (1) [74].

Since the impact of a finite interlayer interaction $J'$ and spin anisotropy $\Delta$ on the critical behavior differ significantly [35, 69], both parameters need to be experimentally determined for a newly synthesized compound to be accurately described as a quasi-2D QHAF. The strength of the antiferromagnetic exchange, $J$, as well as confirmation of lattice and exchange geometry being square rather than rectangular, are of key importance. Secondary considerations include the possible existence of any next-nearest neighbor interactions and spin-canting terms.

For the present case of CuPOF, from the comparison of theoretical modeling to our experimental results of DC susceptibility, specific heat, and high-field magnetization, a leading intralayer exchange constant $J/k_\mathrm{B} = 6.80(5)$ K is consistently determined. All magnetic properties are in excellent agreement with theoretical predictions for the 2D square-lattice spin-1/2 Heisenberg antiferromagnet, see Figs. 2, 4, and 5. The resulting values of $J$, indepen-dently obtained by means of the aforementioned techniques, are in excellent agreement within experimental error. Additionally, the results of DC susceptibility and magnetic specific heat were analyzed in terms of a possible rectangular rather than a square magnetic in-plane structure [57, 58] and are both fully consistent with the square-lattice case.

Table I. Selected quasi-2D spin-1/2 Heisenberg square-lattice antiferromagnets with relevant exchange and anisotropy parameters. The exchange interaction $J$, ordering temperature $T_N$, $g$-factor, the anisotropy field $H_A$, and the saturation field $H_{sat}$ are experimentally determined. The inter- to intralayer coupling ratio $J'/J$ is estimated by use of Eq. (5). The easy-plane anisotropy parameter $\Delta$ is calculated from the out-of-plane static susceptibility minimum by use of Eq. (7). A direct estimate of $\Delta$ from ESR measurements in the ordered state [42] is denoted by [(*)].

| Compound | Ref. | $J/k_B$ (K) | $T_N$ (K) | $k_B T_N/J$ | $J'/J^a$ | $\mu_0 H_A$ (T) | $\mu_0 H_{sat}$ (T) | $H_A/H_{sat}$ | $k_B T_{co}/J$ | $\Delta^b$ |
|---|---|---|---|---|---|---|---|---|---|---|
| [Cu(pz)$_2$(2-HOpy)$_2$](PF$_6$)$_2$ (CuPOF) | this work | 6.8 | 1.38 | 0.203 | $1.4\times10^{-4}$ | 0.36 | 17.6 $\parallel c$ 19.5 $\perp c$ | $1.85\times10^{-2}$ | 0.274 | $0.9\times10^{-2}$ |
| (1) Cu(pz)$_2$(ClO$_4$)$_2$ | [40] | 17.5 | 4.25 | 0.243 | $8.8\times10^{-4}$ | 0.26 | 49 | $5.3\times10^{-3}$ | 0.257 | $4.6\times10^{-3}$ |
| (2) Cu(pz)$_2$(BF$_4$)$_2$ | [40] | 15.3 | 3.8 | 0.248 | $1.1\times10^{-3}$ | 0.25 | 43 | $5.8\times10^{-3}$ | 0.263 | $5.8\times10^{-3}$ |
| (3) [Cu(pz)$_2$(NO$_3$)](PF$_6$) | [40] | 10.8 | 3.05 | 0.282 | $3.3\times10^{-3}$ | 0.007 | 30 | $2.3\times10^{-4}$ | 0.282 | $1.2\times10^{-2}$ |
| (4) [Cu(pz)$_2$(HF$_2$)](PF$_6$) | [42] | 12.8 | 4.38 | 0.342 | $1.4\times10^{-2}$ | n.a. | 33.8 $\parallel c$ 37.5 $\perp c$ | n.a. | n.a. | $3\times10^{-3\,(*)}$ |
| (5) [Cu(pz)$_2$(HF$_2$)](ClO$_4$) | [47] | 7.2 | 1.91 | 0.265 | $1.9\times10^{-3}$ | 0.08 | 20.2 | $4.0\times10^{-3}$ | n.a. | n.a. |
| (6) [Cu(pz)$_2$(pyNO)$_2$](ClO$_4$)$_2$ | [47] | 7.7 | 1.70 | 0.220 | $3.3\times10^{-4}$ | 0.11 | 21.9 | $5.0\times10^{-3}$ | n.a. | n.a. |
| (7) [Cu(pz)$_2$(4-phpyNO)$_2$](ClO$_4$) | [47] | 7.5 | 1.63 | 0.217 | $2.8\times10^{-4}$ | 0.11 | 21.1 | $5.2\times10^{-3}$ | n.a. | n.a. |
| (8) Sr$_2$CuO$_2$Cl$_2$ | [72,73] | 1450 | 255 | 0.176 | $2.4\times10^{-5}$ | 0.7 | 4000$^c$ | $1.8\times10^{-4}$ | 0.221 | $8.3\times10^{-4}$ |

[a] Assuming $\Delta = 0$
[b] Assuming $J' = 0$
[c] Estimated value

In order to further characterize CuPOF, the anisotropy of the electronic $g$-factor is investigated by means of ESR spectroscopy. A very weak in-plane anisotropy was found, with $g_a = 2.073(1)$ and $g_b = 2.066(3)$, in contrast to the out-of-plane $g$-factor $g_c = 2.298(2)$. Overall, the electronic $g$-factor anisotropy evidences a planar-like magnetic structure, with half-filled $d_{x^2-y^2}$ orbitals oriented in the crystallographic $ab$ plane. The integrated ESR intensity, recorded for the out-of-plane field orientation at 9.4 GHz, is in excellent agreement with the bulk magnetic susceptibility, see Fig. 2, and confirms the intralayer antiferromagnetic coupling as $J/k_B = 6.80(5)$ K. Moreover, the linear frequency-field dependence of the ESR spectrum at 1.5 K, observed in the frequency range between 52 and 500 GHz, indicates the absence of a notable energy gap upon approaching the zero-field limit.

The absence of any specific-heat anomaly, see Fig. 5, which would be associated with the transition to long-range order at $T_N = 1.38(2)$ K, as determined by $\mu^+$SR, indicates a high isolation of the magnetic layers and related small change of the residual entropy at $T_N$ [75]. This sets the upper limit of the interlayer coupling as $J'/J < 10^{-2}$ [75]. Due to its pronounced low dimensionality, the thermodynamic properties of CuPOF show a very good agreement with those for an ideal isotropic 2D QHAF, where both, the magnetic susceptibility and the specific heat, exhibit a broad maximum at temperatures of the order of the nearest-neighbor exchange interaction

$J/k_B$. On the other hand, the local-probe techniques $\mu^+$SR and NMR are highly sensitive to the onset of static internal-field components that associated with long-range order. $\mu^+$SR was successfully used to determine the long-range ordering temperature $T_N$ for a wide range of quasi-2D square-lattice quantum Heisenberg antiferromagnets [44, 46, 47, 63].

In the case of CuPOF, the temperature evolution of the asymmetry parameter, precession frequency, and the relaxation rate of the oscillatory component of the $\mu^+$SR asymmetry relaxation function Eq. (3) revealed a zero-field transition to long-range order at $T_N = 1.38(2)$ K, see Fig. 6. Thus, the ratio $k_B T_N/J$ is found to be 0.203 for CuPOF, which is the smallest among all yet-characterized molecular-based materials with a magnetic lattice of $Cu^{2+}$ ions, compare Table I. By use of the empirical formula Eq. (5), the inter- to intralayer exchange ratio is obtained as $J'/J = 1.4 \times 10^{-4}$, with $J' \simeq 1$ mK. Since strictly isotropic exchange interactions, $\Delta = 0$, are assumed in the derivation of Eq. (5), $J'/J = 1.4 \times 10^{-4}$ sets an upper limit to the effective interlayer coupling. The dipolar interaction between nearest-neighbor $Cu^{2+}$ ions in adjacent layers, with a Cu–Cu distance of 13.097 Å, is estimated as about 1 mK. Therefore, the interlayer interaction likely stems from dipole-dipole coupling, rather than superexchange via the interlayer molecules. A planar magnetic structure with highly isolated layers is further supported by our density functional theory (DFT) calculations [76, 77], see SM. Both the interlayer interactions and the next-nearest-neighbor intralayer interactions are negligible in comparison to the antiferromagnetic in-plane coupling, mediated by the pyrazine molecules.

As discussed above, the anisotropy parameter $\Delta$ can be estimated from the low-temperature magnetization or low-field susceptibility measurements [26, 40, 47, 69, 73, 78]. From the anisotropy field $\mu_0 H_A = 0.36(1)$ T, see inset of Fig. 4(b), the exchange anisotropy in CuPOF is evaluated as $\Delta = H_A / H_{sat} = 1.85(5) \times 10^{-2}$. This results in the largest yet found ratio $H_A / H_{sat}$ for Cu-pyrazine based layered materials, compare Table I. A slightly smaller anisotropy parameter is estimated from the out-of-plane DC susceptibility at 0.1 T, see inset of Fig. 3. Employing the empirical Eq. (7), derived for the case of $J' = 0$, the anisotropy parameter $\Delta = 0.9(2) \times 10^{-2}$ is determined. The qualitative behavior of the field-dependent magnetization (Fig. 4) and temperature-dependent DC susceptibility (Fig. 3) are in very good agreement with calculations for a 2D QHAF with an $XY$ anisotropy of 1–2% of the intralayer coupling [26, 69], and resembles the characteristic behavior of previously studied Cu-pyrazine based compounds [40, 47, 73].

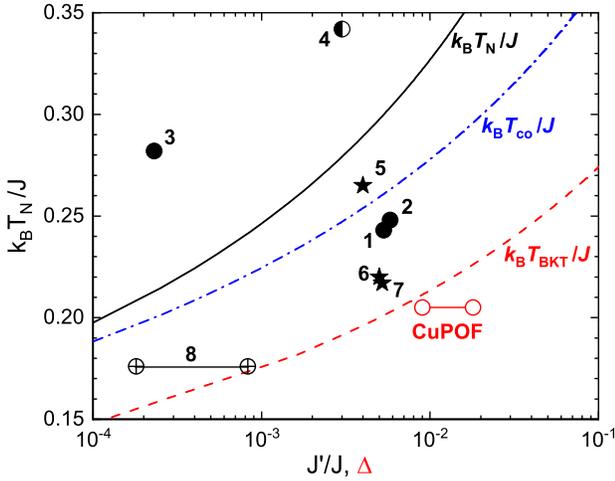

Figure 8. Normalized Néel temperature as a function of $J'/J$ (black solid line), as well as the normalized crossover (blue dash-dotted line) and BKT transition temperatures (red dashed line) as functions of the anisotropy parameter $\Delta$. The normalized Néel temperature, $k_B T_N / J$, as a function of intrinsic easy-plane anisotropy, $\Delta$, for various Cu-pyrazine 2D QHAFs (○, ●, ◐, ★) and Sr$_2$CuO$_2$Cl$_2$ (⊕) [72,73]. The values indicated by ○ are for CuPOF, ● are from Ref. [40], ◐ is from Ref. [42], and ★ are from Ref. [47], respectively. The numbers denote 2D QHAF compounds according to the labeling in Table I. The two distinct values for CuPOF and Sr$_2$CuO$_2$Cl$_2$ denote slightly different values of $\Delta$, determined from the DC susceptibility and low-field magnetization measurements, respectively.

In order to evaluate the relevant perturbations with respect to the ideal 2D QHAF and to shed light on the driving mechanism of the long-range order in CuPOF, the normalized Néel temperature, $k_B T_N / J$, is shown as a function of the evaluated spin anisotropy $\Delta$ in Fig. 8, and compared to other molecular-based quasi-2D QHAFs. Where possible, the parameter $\Delta$ was determined from experimental values of $H_A / H_{sat}$ for the compounds labeled by black solid circles and stars. The two values for the inorganic compound Sr$_2$CuO$_2$Cl$_2$ are based on $H_A / H_{sat} = 1.8 \times 10^{-4}$ [19] and the minimum of the out-of-plane DC susceptibility $k_B T_{co} / J = 0.221$ [73], which corresponds to $\Delta = 8.3 \times 10^{-4}$, see Table I. The experimental values are compared to the theoretical expectation of the BKT transition temperature for a weakly anisotropic 2D QHAF, described by the empirical function Eq. (6).

For highly isolated quasi-2D materials, such as Sr$_2$CuO$_2$Cl$_2$ with $J'/J = 3 \times 10^{-5}$, the experimentally observed value of $k_B T_N / J$ is very close to the prediction of $k_B T_{BKT} / J$ (red dashed line in Fig. 8). Although the transitions to long-range and topological order at $T_N$ and $T_{BKT}$, respectively, are of different nature, it was argued that, due to the exceptional low dimensionality of Sr$_2$CuO$_2$Cl$_2$, the long-range order is triggered by the inceptive BKT-type topological transition at $T_{BKT} \lesssim T_N$ [25, 26]. Due to the finite interlayer coupling, a Néel-type antiferromagnetic order is stabilized at $T_N$, before the topological transition is completed. A very similar scenario is proposed for CuPOF, motivated by the close agreement between the experimentally obtained value of $T_N$, and the theoretically predicted value of $T_{BKT}$. Moreover, the larger ratio $k_B T_N / J = 0.203$ as compared to $k_B T_N / J = 0.176$ for Sr$_2$CuO$_2$Cl$_2$, is attributed *only* to the stronger intrinsic spin anisotropy in CuPOF.

Furthermore, the field-driven increase of the crossover temperature $T_{co}$, observed by DC susceptibility, see Fig. 3, reveals the corresponding increase of the effective spin anisotropy. This field-induced spin anisotropy is evidenced as well by the strong increase of the ordering temperature $T_N$ in applied magnetic field, as found by [1]H-NMR spectroscopy, see Fig. 7. This strong increase of $T_N$ is attributed to the exceptional two-dimensionality of CuPOF, as compared to other less-isolated quasi-2D Cu-pyrazine-layered compounds, for which weaker field-induced changes of $T_N$ were reported [42, 45, 79, 80]. The splitting of the [1]H-NMR spectrum below the transition to long-range order is a clear signature of commensurate antiferromagnetic order.

In conclusion, we present a comprehensive characterization of the newly synthesized molecular-based compound [Cu(pz)$_2$(2-HOpy)$_2$](PF$_6$)$_2$ (CuPOF). Employing bulk magnetometry, specific heat, density functional theory calculations, ESR, $\mu^+$SR, and NMR spectroscopy, CuPOF is characterized as an excellent realization of the 2D square-lattice spin-1/2 Heisenberg model with a moderate nearest-neighbor exchange interaction of $J/k_B = 6.80(5)$ K, and well-separated magnetic layers. The intralayer interaction is about four orders of magnitude larger than the estimated upper limit on the interlayer interaction, $J' \simeq 1$ mK. A weak intrinsic easy-plane anisotropy, revealed by bulk magnetometry, yields a temperature-driven crossover of the spin correlations from isotropic Heisenberg to anisotropic $XY$-type behavior, and constitutes the driving mechanism of a transition to magnetic long-range order at $T_N = 1.38(2)$ K, as revealed by $\mu^+$SR spectroscopy. The application of a magnetic field normal to the easy-plane yields a field-driven increase of the magnetic anisotropy, as shown by the evolution of the crossover temperature $T_{co}$ in the DC susceptibility data. The application of magnetic fields of several tesla leads to a strong increase of $T_N$, as revealed by $^1$H-NMR spectroscopy, in agreement with the pronounced two dimensionality of the magnetic lattice in CuPOF. As an outlook, our comprehensive characterization of $[Cu(pz)_2(2\text{-HOpy})_2](PF_6)_2$ as a clean realization of a 2D square-lattice spin-1/2 Heisenberg antiferromagnet with a moderate intralayer coupling and highly isolated magnetic layers calls for further studies of the field-induced effects on the anisotropy of the magnetic correlations [37], in particular by scattering and local-probe techniques.

Data presented in this paper resulting from the UK effort will be made available [81].

**ACKNOWLEDGMENTS**

Christoph Klausnitzer from the Max Planck Institute for Chemical Physics of Solids is acknowledged for his strong technical support, and in memory of his lifetime achievements. We thank Paul Gape for experimental assistance with the muon experiment and for discussion of the data analysis. We acknowledge the support of HLD at HZDR, member of the European Magnetic Field Laboratory (EMFL). We acknowledge the support from the Deutsche Forschungsgemeinschaft (DFG) through SFB 1143 and the Würzburg-Dresden Cluster of Excellence on Complexity and Topology in Quantum Matter — ct.qmat (EXC 2147, project No. 390858490). Work at the National High Magnetic Field Laboratory is supported by NSF Cooperative Agreements No. DMR-1157490 and No. DMR-1644779, the State of Florida, the US DOE, and the DOE Basic Energy Science Field Work Project Science in 100 T. This project has received funding from the European Research Council (ERC) under the European Union's Horizon 2020 research and innovation program (Grant Agreement No. 681260). Part of this work was carried out at the STFC-ISIS facility and we are grateful for provision of beamtime. This work is supported by EPSRC (UK). S.A.Z. and A.N.P. acknowledge the support of DFG through ZV 6/2-2. I.H. and R. S. were supported by the European Regional Development Fund (Grant No. TK134), and by the Estonian Research Council (PRG4, IUT23-7).

[1] L. J. de Jongh and A. R. Miedema, Adv. Phys. **23**, 1 (1974).
[2] L. Onsager, Phys. Rev. **65**, 117 (1944).
[3] H.-Q. Ding, J. Phys.: Condens. Matter **2**, 7979 (1990).
[4] N. D. Mermin and H. Wagner, Phys. Rev. Lett. **17**, 1133 (1966).
[5] V. L. Berezinskii, Sov. Phys. JETP **32**, 493 (1971).
[6] J. M. Kosterlitz and D. J. Thouless, J. Phys. C: Solid State Phys. **6**, 1181 (1973).
[7] H.-Q. Ding and M. S. Makivić, Phys. Rev. Lett. **64**, 1449 (1990).
[8] J. G. Bednorz and K. A. Müller, Z. Phys. B: Condens. Matter **64**, 189 (1986).
[9] P. W. Anderson, Science **235**, 1196 (1987).
[10] Z. Z. Sheng and A. M. Hermann, Nature **332**, 55 (1988).
[11] A. Schilling, M. Cantoni, J. D. Guo, and H. R. Ott, Nature **363**, 56 (1993).
[12] C. W. Chu, L. Gao, F. Chen, Z. J. Huang, R. L. Meng, and Y. Y. Xue, Nature **365**, 323 (1993).
[13] S. N. Putilin, E. V. Antipov, O. Chmaissem, and M. Marezio, Nature **362**, 226 (1993).
[14] A. J. Leggett, Nat. Phys. **2**, 134 (2006).
[15] J. R. Kirtley, C. C. Tsuei, Ariando, C. J. M. Verwijs, S. Harkema, and H. Hilgenkamp, Nat. Phys. **2**, 190 (2006).
[16] S. Graser, P. J. Hirschfeld, T. Kopp, R. Gutser, B. M. Andersen, and J. Mannhart, Nat. Phys. **6**, 609 (2010).
[17] S. Chakravarty, B. I. Halperin, and D. R. Nelson, Phys. Rev. Lett. **60**, 1057 (1988).
[18] B. Keimer, N. Belk, R. J. Birgeneau, A. Cassanho, C. Y. Chen, M. Greven, M. A. Kastner, A. Aharony, Y. Endoh, R. W. Erwin, and G. Shirane, Phys. Rev. B **46**, 14034 (1992).
[19] M. Greven, R. J. Birgeneau, Y. Endoh, M. A. Kastner, M. Matsuda, and G. Shirane, Z. Phys. B: Condens. Matter **96**, 465 (1995).
[20] M. Greven, R. J. Birgeneau, Y. Endoh, M. A. Kastner, B. Keimer, M. Matsuda, G. Shirane, and T. R. Thurston, Phys. Rev. Lett. **72**, 1096 (1994).
[21] B.-G. Liu, Phys. Rev. B **41**, 9563 (1990).
[22] N. Majlis, S. Selzer, and G. C. Strinati, Phys. Rev. B **45**, 7872 (1992).
[23] N. Majlis, S. Selzer, and G. C. Strinati, Phys. Rev. B **48**, 957 (1993).
[24] L. Siurakshina, D. Ihle, and R. Hayn, Phys. Rev. B **61**, 14601 (2000).
[25] H.-Q. Ding, Phys. Rev. Lett. **68**, 1927 (1992).
[26] A. Cuccoli, T. Roscilde, R. Vaia, and P. Verrucchi, Phys. Rev. Lett. **90**, 167205 (2003).
[27] T. Yildirim, A. B. Harris, O. Entin-Wohlman, and A. Aharony, Phys. Rev. Lett. **72**, 3710 (1994).
[28] D. J. Bishop and J. D. Reppy, Phys. Rev. Lett. **40**, 1727 (1978).
[29] H. J. F. Knops, Phys. Rev. Lett. **39**, 766 (1977).
[30] B. I. Halperin and D. R. Nelson, Phys. Rev. Lett. **41**, 121 (1978).
[31] D. J. Resnick, J. C. Garland, J. T. Boyd, S. Shoemaker, and R. S. Newrock, Phys. Rev. Lett. **47**, 1542 (1981).

[32] A. I. Safonov, S. A. Vasilyev, I. S. Yasnikov, I. I. Lukashevich, and S. Jaakkola, Phys. Rev. Lett. **81**, 4545 (1998).

[33] Z. Hadzibabic, P. Krüger, M. Cheneau, B. Battelier, and J. Dalibard, Nature **441**, 1118 (2006).

[34] K. Kim, S. Y. Lim, J.-U. Lee, S. Lee, T. Y. Kim, K. Park, G. S. Jeon, C.-H. Park, J.-G. Park, and H. Cheong, Nat. Commun. **10**, 1 (2019).

[35] C. Yasuda, S. Todo, K. Hukushima, F. Alet, M. Keller, M. Troyer, and H. Takayama, Phys. Rev. Lett. **94**, 217201 (2005).

[36] D. P. Landau and K. Binder, Phys. Rev. B **24**, 1391 (1981).

[37] A. Cuccoli, T. Roscilde, R. Vaia, and P. Verrucchi, Phys. Rev. B **68**, 060402(R) (2003).

[38] F. M. Woodward, P. J. Gibson, G. B. Jameson, C. P. Landee, M. M. Turnbull, and R. D. Willett, Inorg. Chem. **46**, 4256 (2007).

[39] P. A. Goddard, J. Singleton, P. Sengupta, R. D. McDonald, T. Lancaster, S. J. Blundell, F. L. Pratt, S. Cox, N. Harrison, J. L. Manson, H. I. Southerland, and J. A. Schlueter, New J. Phys. **10**, 083025 (2008).

[40] F. Xiao, F. M. Woodward, C. P. Landee, M. M. Turnbull, C. Mielke, N. Harrison, T. Lancaster, S. J. Blundell, P. J. Baker, P. Babkevich, and F. L. Pratt, Phys. Rev. B **79**, 134412 (2009).

[41] J. L. Manson, K. H. Stone, H. I. Southerland, T. Lancaster, A. J. Steele, S. J. Blundell, F. L. Pratt, P. J. Baker, R. D. McDonald, P. Sengupta, J. Singleton, P. A. Goddard, C. Lee, M.-H. Whangbo, M. M. Warter, C. H. Mielke, and P. W. Stephens, J. Am. Chem. Soc. **131**, 4590 (2009).

[42] E. Čižmár, S. A. Zvyagin, R. Beyer, M. Uhlarz, M. Ozerov, Y. Skourski, J. L. Manson, J. A. Schlueter, and J. Wosnitza, Phys. Rev. B **81**, 064422 (2010).

[43] N. S. Headings, S. M. Hayden, R. Coldea, and T. G. Perring, Phys. Rev. Lett. **105**, 247001 (2010).

[44] A. J. Steele, T. Lancaster, S. J. Blundell, P. J. Baker, F. L. Pratt, C. Baines, M. M. Conner, H. I. Southerland, J. L. Manson, and J. A. Schlueter, Phys. Rev. B **84**, 064412 (2011).

[45] Y. Kohama, M. Jaime, O. E. Ayala-Valenzuela, R. D. McDonald, E. D. Mun, J. F. Corbey, and J. L. Manson, Phys. Rev. B **84**, 184402 (2011).

[46] P. A. Goddard, J. L. Manson, J. Singleton, I. Franke, T. Lancaster, A. J. Steele, S. J. Blundell, C. Baines, F. L. Pratt, R. D. McDonald, O. E. Ayala-Valenzuela, J. F. Corbey, H. I. Southerland, P. Sengupta, and J. A. Schlueter, Phys. Rev. Lett. **108**, 077208 (2012).

[47] P. A. Goddard, J. Singleton, I. Franke, J. S. Möller, T. Lancaster, A. J. Steele, C. V. Topping, S. J. Blundell, F. L. Pratt, C. Baines, J. Bendix, R. D. McDonald, J. Brambleby, M. R. Lees, S. H. Lapidus, P. W. Stephens, B. W. Twamley, M. M. Conner, K. Funk, J. F. Corbey, H. E. Tran, J. A. Schlueter, and J. L. Manson, Phys. Rev. B **93**, 094430 (2016).

[48] V. Selmani, C. P. Landee, M. M. Turnbull, J. L. Wikaira, and F. Xiao, Inorg. Chem. Commun. **13**, 1399 (2010).

[49] See Supplemental Material for the additional information about the material synthesis, crystal structure, magnetometry, ESR spectroscopy, specific heat, MAS NMR, and DFT calculations results, as well as description of the μSR setup, which includes Refs. [82–95].

[50] CrysAlisPro, Oxford Diffraction/Agilent Technologies UK Ltd, Yarnton, England.

[51] SCALE3 ABSPACK, version: 1.0.4, gui: 1.03; an oxford diffraction program; Oxford Diffraction Ltd: Abingdon, UK (2005).

[52] G. M. Sheldrick, Acta Cryst. A **64**, 112 (2008).

[53] G. M. Sheldrick, Acta Cryst. C **71**, 3 (2015).

[54] S. Stoll and A. Schweiger, J. Magn. Reson. **178**, 42 (2006).

[55] S. Zvyagin, J. Krzystek, P. van Loosdrecht, G. Dhalenne, and A. Revcolevschi, Phys. B (Amsterdam, Neth.) **346-347**, 1 (2004).

[56] F. M. Woodward, A. S. Albrecht, C. M. Wynn, C. P. Landee, and M. M. Turnbull, Phys. Rev. B **65**, 144412 (2002).

[57] B. C. Keith, C. P. Landee, T. Valleau, M. M. Turnbull, and N. Harrison, Phys. Rev. B **84**, 104442 (2011).

[58] B. C. Keith, C. P. Landee, T. Valleau, M. M. Turnbull, and N. Harrison, Phys. Rev. B **84**, 229901(E) (2011).

[59] A. Bencini and D. Gatteschi, *EPR of Exchange Coupled Systems* (Dover Publications, Mineola, 2012).

[60] P. M. Richards and M. B. Salamon, Phys. Rev. B **9**, 32 (1974).

[61] The average values of the g-factors from two independent ESR measurements on the single-crystalline CuPOF samples are presented here; for details see SM.

[62] M. Dumm, A. Loidl, B. W. Fravel, K. P. Starkey, L. K. Montgomery, and M. Dressel, Phys. Rev. B **61**, 511 (2000).

[63] T. Lancaster, S. J. Blundell, M. L. Brooks, P. J. Baker, F. L. Pratt, J. L. Manson, M. M. Conner, F. Xiao, C. P. Landee, F. A. Chaves, S. Soriano, M. A. Novak, T. P. Papageorgiou, A. D. Bianchi, T. Herrmannsdörfer, J. Wosnitza, and J. A. Schlueter, Phys. Rev. B **75**, 094421 (2007).

[64] A. Sandvik and C. P. Landee, unpublished.

[65] S. J. Blundell, Contemp. Phys. **40**, 175 (1999).

[66] T. Lancaster, S. J. Blundell, P. J. Baker, M. L. Brooks, W. Hayes, F. L. Pratt, J. L. Manson, M. M. Conner, and J. A. Schlueter, Phys. Rev. Lett. **99**, 267601 (2007).

[67] A. Samoson, T. Tuherm, J. Past, A. Reinhold, T. Anupold, and I. Heinmaa, Top. Curr. Chem. **246**, 15 (2005).

[68] A. W. Sandvik, Phys. Rev. Lett. **80**, 5196 (1998).

[69] A. Cuccoli, T. Roscilde, V. Tognetti, R. Vaia, and P. Verrucchi, Phys. Rev. B **67**, 104414 (2003).

[70] S. Chakravarty, B. I. Halperin, and D. R. Nelson, Phys. Rev. B **39**, 2344 (1989).

[71] A. Cuccoli, T. Roscilde, V. Tognetti, R. Vaia, and P. Verrucchi, Eur. Phys. J. B **20**, 55 (2001).

[72] B. J. Suh, F. Borsa, L. L. Miller, M. Corti, D. C. Johnston, and D. R. Torgeson, Phys. Rev. Lett. **75**, 2212 (1995).

[73] D. Vaknin, S. K. Sinha, C. Stassis, L. L. Miller, and D. C. Johnston, Phys. Rev. B **41**, 1926 (1990).

[74] L. J. de Jongh (ed.), *Magnetic Properties of Layered Transition Metal Compounds* (Kluwer, Dordrecht, 1990).

[75] P. Sengupta, A. W. Sandvik, and R. R. P. Singh, Phys. Rev. B **68**, 094423 (2003).

[76] S. Vela, A. Sopena, J. Ribas-Arino, J. J. Novoa, and M. Deumal, Chem. Eur. J. **20**, 7083 (2014).

[77] S. Vela, J. Jornet-Somoza, M. M. Turnbull, R. Feyerherm, J. J. Novoa, and M. Deumal, Inorg. Chem. **52**, 12923 (2013).

[78] M. Matsuura, K. Gilijamse, J. E. W. Sterkenburg, and D. J. Breed, Phys. Lett. A **33**, 363 (1970).

[79] P. Sengupta, C. D. Batista, R. D. McDonald, S. Cox, J. Singleton, L. Huang, T. P. Papageorgiou, O. Ignatchik, T. Herrmannsdörfer, J. L. Manson, J. A. Schlueter, K. A. Funk, and J. Wosnitza, Phys. Rev. B **79**, 060409(R) (2009).

[80] N. A. Fortune, S. T. Hannahs, C. P. Landee, M. M. Turnbull, and F. Xiao, J. Phys. Conf. Ser. **568**, 042004 (2014).

[81] https://wrap.warwick.ac.uk/139957

[82] E. A. Turov, *Physical properties of magnetically ordered crystals* (New York: Acad. Press, 1965).

[83] L. C. M. Gorkom, J. M. Hook, M. B. Logan, J. V. Hanna, and R. E. Wasylishen, Magn. Reson. Chem. **33**, 791 (1995).

[84] E. R. Andrew, M. Firth, A. Jasinski, and P. Randall, Phys. Lett. A **31**, 446 (1970).

[85] E. C. Alyea, J. Malito, and J. H. Nelson, Inorg. Chem. **26**, 4294 (1987).

[86] M. Alla and E. Lippmaa, Chem. Phys. Lett. **87** (1982).

[87] L. Noodleman and E. R. Davidson, Chem. Phys. **109**, 131 (1986).

[88] L. Noodleman, J. Chem. Phys. **74**, 5737 (1981).

[89] T. Soda, Y. Kitagawa, T. Onishi, Y. Takano, Y. Shigeta, H. Nagao, Y. Yoshioka, and K. Yamaguchi, Chem. Phys. Lett. **319**, 223 (2000).

[90] M. J. Frisch *et al.*, *Gaussian 09* (Gaussian, Inc., Wallingford CT, 2009).

[91] A. D. Becke, J. Chem. Phys. **104**, 1040 (1996).

[92] A. D. Becke, Phys. Rev. A **38**, 3098 (1988).

[93] C. Lee, W. Yang, and R. G. Parr, Phys. Rev. B **37**, 785 (1988).

[94] J. J. Novoa, M. Deumal, and J. Jornet-Somoza, Chem. Soc. Rev. **40**, 3182 (2011).

[95] M. Deumal, M. J. Bearpark, J. J. Novoa, and M. A. Robb, J. Phys. Chem. A **106**, 1299 (2002).

# Extremely well isolated 2D spin-1/2 antiferromagnetic Heisenberg layers with small exchange coupling in the molecular-based magnet CuPOF:
## Supplemental Material

D. Opherden,[1, 2, *] N. Nizar,[3] K. Richardson,[3] J. C. Monroe,[4] M. M. Turnbull,[4] M. Polson,[5] S. Vela,[6]
W. J. A. Blackmore,[7] P. A. Goddard,[7] J. Singleton,[8] E. S. Choi,[9] F. Xiao,[10] R. C. Williams,[10]
T. Lancaster,[10] F. L. Pratt,[11] S. J. Blundell,[12] Y. Skourski,[1] M. Uhlarz,[1] A. N. Ponomaryov,[1]
S. A. Zvyagin,[1] J. Wosnitza,[1, 2] M. Baenitz,[13] I. Heinmaa,[14] R. Stern,[14] H. Kühne,[1, †] and C. P. Landee[3, ‡]

[1]*Hochfeld-Magnetlabor Dresden (HLD-EMFL) and Würzburg-Dresden Cluster of Excellence ct.qmat,
Helmholtz-Zentrum Dresden-Rossendorf, 01328 Dresden, Germany*
[2]*Institut für Festkörper- und Materialphysik, TU Dresden, 01062 Dresden, Germany*
[3]*Department of Physics, Clark University, Worcester, MA 01610, USA*
[4]*Carlson School of Chemistry, Clark University, Worcester, MA 01610, USA*
[5]*Department of Chemistry, University of Canterbury, Private Bag 4800, Christchurch, New Zealand*
[6]*Laboratoire de Chimie Quantique, UMR 7177, CNRS-Universitéde Strasbourg,
4, rue Blaise Pascal, F-67000 Strasbourg, France*
[7]*Department of Physics, University of Warwick, Gibbet Hill Road, Coventry, CV4 7AL, UK*
[8]*National High Magnetic Field Laboratory, Los Alamos National Laboratory, Los Alamos, New Mexico 87545, USA*
[9]*National High Magnetic Field Laboratory, Florida State University, Tallahassee, FL 32310, USA*
[10]*Durham University, Centre for Materials Physics, South Road, Durham DH1 3LE, UK*
[11]*ISIS Facility, Rutherford Appleton Laboratory, Chilton, Didcot, Oxford OX11 0QX, UK*
[12]*Clarendon Laboratory, Department of Physics, University of Oxford, Park Road, Oxford OX1 3PU, UK*
[13]*Max Planck Institute for Chemical Physics of Solids, 01187 Dresden, Germany*
[14]*National Institute of Chemical Physics and Biophysics, Akadeemia tee 23, 12618 Tallinn, Estonia*
(Dated: August 31, 2020)

*Synthesis and Crystal Structure.* Copper(II) bromide (0.58 g, 2.5 mmol) and silver hexafluorophosphate (1.261 g, 5.0 mmol) were separately dissolved in 5 ml of methanol each. The copper-bromide solution was added to the other solution while stirring; an immediate precipitate resulted. The combined solution was stirred for thirty minutes, then filtered. The precipitate yielded a dry weight of 0.820 g, equivalent to 4.36 mmol of AgBr (87% yield). Pyrazine (0.40 g, 5.0 mmol) and 2-hydroxypyridine (0.475 g, 5.0 mmol) were dissolved separately in 5 ml of MeOH. These solutions were combined and added dropwise to the stirred filtrate; the green cupric solution gradually changed to an olive-green color but no precipitate appeared. The solution was stirred for one hour, filtered, then partially covered and set to evaporate. Within several days, flat green crystals appeared. The product was collected by vacuum filtration, rinsed once with MeOH, then dried under vacuum. The crystals obtained were pale green, pleochroic plates, several mm on a side. Under one orientation of polarized light, the crystals are emerald green, and sky blue along a perpendicular orientation. The crystals extinguish between crossed polarizers when the growth edges are parallel to the axes of polarization. Total product (0.504 g, 0.716 mmol) for a yield of 28.6%. The infrared spectrum by the attenuated total reflection (IR (ATR)) and the results of the elemental analysis (CHN) appear in the following. IR (ATR): 3411w, 3134w, 1646s, 1595s, 1543m, 1455m, 1425s, 1401m, 1367w, 1258w, 1231w, 1158m, 1170w, 1155w, 1120m, 1094w, 1086w, 1066m, 1031s, 997m, 832s, 822s, 778w, 765s, 738m, 709s, 555s, 532w 5507m cm$^{-1}$. (CHN) calculated (experimental): C 30.7 (30.9, 30.9), H 2.58 (2.63, 2.49), N 11.94 (11.76, 11.90).

The local coordination sphere of the $Cu^{2+}$ ion in $[Cu(pyz)_2(2\text{-HOpy})_2](PF_6)_2$ (CuPOF) is presented in Fig. S1. The copper ion and the four pyrazine nitrogen atoms lie in the $ab$ plane, whereas the coordinated 2-pyridone molecule is nearly normal to the plane. Symmetry equivalent atoms A are generated via a mirror plane perpendicular to the $a$ axis, whereas symmetry equivalent atoms B are generated by a two-fold rotation axis parallel to $b$.

The crystal data and structure-refinement parameters for $[Cu(pyz)_2(2\text{-HOpy})_2](PF_6)_2$, as well as selected bond lengths and angles are presented in Tables S1 and S2, respectively.

*Magnetometry.* The single-crystal susceptibilities of $[Cu(pyz)_2(2\text{-HOpy})_2](PF_6)_2$ between 1.8 and 12 K are shown in Fig. S2, where they are plotted as the ratios $\chi_i/C_i$ to demonstrate the equality of the temperature dependence for the out-of-plane and in-plane directions $c$ and $ab$, respectively, where $C_i$ is the corresponding Curie constant. $C_c$ and $C_{ab}$ are 0.496 and 0.402 emuG$^{-1}$mol$^{-1}$K, respectively. This equivalence demonstrates that the ideal Heisenberg model is appropriate to describe the data in this temperature range.

————
* Corresponding author. E-mail: d.dmytriieva@hzdr.de
† Corresponding author. E-mail: h.kuehne@hzdr.de
‡ Corresponding author. E-mail: CLandee@clarku.edu

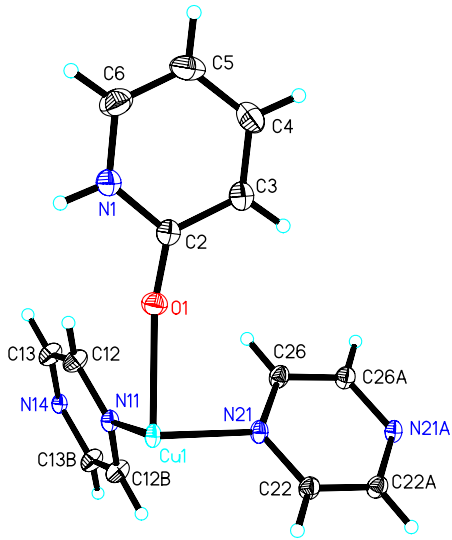

FIG. S1. The local coordination sphere of the magnetic $Cu^{2+}$ ions in CuPOF.

TABLE S1. Crystal data and structure-refinement parameters for $[Cu(pyz)_2(2\text{-HOpy})_2](PF_6)_2$.

| | |
|---|---|
| Empirical formula | $C_{18}H_{18}CuF_{12}N_6O_2P_2$ |
| Formula weight | 703.86 g/mol |
| Temperature | 120(2) K |
| Wavelength | 1.54184 Å |
| Crystal system | Orthorhombic |
| Space group | $Cmca$ |
| Unit-cell dimensions | $a = 13.75960(18)$ Å |
| | $b = 13.75212(19)$ Å |
| | $c = 25.7865(4)$ Å |
| | $\alpha = \beta = \gamma = 90°$ |
| Volume | 4879.42(12) Å$^3$ |
| $Z$ | 8 |
| Density (calculated) | 1.916 Mg/m$^3$ |
| Absorption coefficient | 3.686 mm$^{-1}$ |
| F(000) | 2808 |
| $\theta$ range for data collection | 3.428° to 76.330° |
| Index ranges | $-10 \leq h \leq 17$ |
| | $-17 \leq k \leq 17$ |
| | $-32 \leq l \leq 32$ |
| Reflections collected | 23509 |
| Independent reflections | 2683 [$R$(int) = 0.0401] |
| Completeness to $\theta = 67.684°$ | 100.0% |
| Absorption correction | Gaussian |
| Max. and min. transmission | 0.851 and 0.267 |
| Refinement method | Full-matrix least-squares on F2 |
| Data/restraints/parameters | 2683/0/208 |
| Goodness-of-fit on F2 | 1.060 |
| Final $R$ indices [$I > 2\sigma(I)$] | $R_1 = 0.0368$, $wR_2 = 0.1020$ |
| $R$ indices (all data) | $R_1 = 0.0387$, $wR_2 = 0.1036$ |
| Largest diff. peak and hole | 1.068 and -0.563 e·Å$^{-3}$ |

In order to obtain the absolute values of the magnetization recorded in the pulsed-field experiments, the data were directly compared with the magnetization recorded for a CuPOF single crystal at 4 K for $H \parallel c$ (at 2 K for $H \perp c$) and up to 14 T, using a vibrating sample magnetometer (VSM) in a superconducting magnet and $^4$He cryostat. These results are presented in Fig. S3.

Additional measurements of the magnetization of a polycrystalline sample in DC fields up to 35 T were done using a VSM magnetometer at the National High Magnetic Field Laboratory in Tallahassee, as well as in pulsed fields up to 25 T at the NHMFL facility in Los Alamos. Representative examples of these data are presented in Fig. S4. The results are fully consistent with the data from the HLD (see Fig. 3 in the main text).

*Electron Spin Resonance.* The room-temperature ESR powder spectrum of $[Cu(pyz)_2(2\text{-HOpy})_2](PF_6)_2$, measured at 9.8 GHz, is shown in Fig. S5. A least-squares fit to the spectrum performed with EasySpin is shown as the red line. The fit to the spectrum indicates a slightly rhombic $g$-tensor with $g_a = 2.071$, $g_b = 2.056$, and $g_c = 2.322$. The lineshape is primarily Lorentzian with a smaller Gaussian contribution, as indicated by the fit parameters 1.207 mT (Lorentzian contribution) and 0.524 mT (Gaussian contribution). The parameters of the chosen pseudo-Voigt function represent coefficients in the linear combination of Lorentzian and Gaussian contributions to the lineshape [1]. The mostly Lorentzian lineshape indicates a dominant superexchange interaction [2]. A small value of $G_{strain} = 0.027$ mT was permitted in fitting the lineshape contribution of $g_c$, where

$G_{strain}$ accounts for anisotropic broadening due to a local distribution of the $g$-factor, by adding a Gaussian envelope to the line [1]. The average $g$-factor determined from the powder ESR spectrum, $\langle g_{pow} \rangle = 2.150$, agrees well with the values determined from the single crystal measurements, $\langle g_{SC} \rangle = 2.146(3)$, as well as from the magnetic DC susceptibility, $\langle g_\chi \rangle = 2.167$.

The anisotropy of the room-temperature ESR spectrum of a single-crystalline CuPOF sample was investigated at 9.8 GHz. A single ESR line of nearly Lorentzian lineshape was observed for each field orientation. As a representative example, the ESR spectrum obtained for a field along the crystallographic $a$ axis is shown in Fig. S6.

The angular dependences of the $g$-factor in the $ac$, $bc$, and $ab$ planes are presented in Fig. S7, resulting in $g_a = 2.074(1)$, $g_b = 2.068(2)$, and $g_c = 2.300(1)$, see Table S3. The $g$-factors along the three crystallographic axes are determined from fits with a $\cos^2(\theta)$-type angular dependence, shown by the red solid lines in Fig. S7. Additional measurements of the angular-dependent ESR spectra of a single-crystalline sample of CuPOF were per-formed at the Dresden High Magnetic Field Laboratory. The anisotropy of the $g$-factor obtained from these measurements is in very good agreement with those determined from the measurements at Clark University. A comparison of all experimentally determined $g$-factors and linewidths is presented in Table S3.

TABLE S2. Selected bond lengths and angles for $[Cu(pyz)_2(2\text{-}HOpy)_2](PF_6)_2$. Symmetry transformations used to generate equivalent atoms: #1 $(-x+1/2, y+0, -z+1/2)$, #2 $(x, y-1/2, -z+1/2)$, #3 $(x, y+1/2, -z+1/2)$.

| | |
|---|---|
| Cu1–N11 | 2.049(2) Å |
| Cu1–N21 | 2.0508(16) Å |
| Cu1–O1 | 2.2851(14) Å |
| N14–Cu1(#3) | 2.057(2) Å |
| N11–Cu1–N21 | 90.08(4)° |
| N21–Cu1–N21(#1) | 179.84(7)° |
| N11–Cu1–N14(#2) | 180.0° |
| N21–Cu1–N14(#2) | 89.92(4)° |
| N11–Cu1–O1 | 89.01(3)° |
| N21–Cu1–O1 | 90.23(5)° |
| N14(#2)–Cu1–O1 | 90.99(3)° |
| N21–Cu1–O1 | 89.77(5)° |
| O1–Cu1–O1(#1) | 178.02(7)° |
| C2–O1–Cu1 | 168.01(12)° |

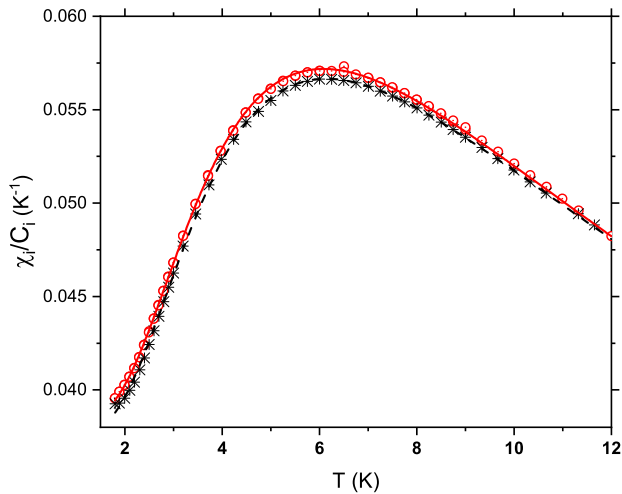

FIG. S2. Temperature dependence of the single-crystal susceptibilities with the field normal to the layer ($\chi_c$, black stars) and within the layer ($\chi_{ab}$, red circles). The data are presented as the ratios $\chi_i/C_i$ to demonstrate the equivalence of the temperature dependencies, where $C_i$ are the corresponding Curie constants. The red solid line and black dashed line represent the best fits of the 2D QHAF model calculations to the magnetic susceptibilities $\chi_c$ and $\chi_{ab}$ with $J/k_B = 6.75(2)$ and 6.71(1) K, respectively.

The angular dependences $\Delta H_{pp}(\theta)$ of the room-

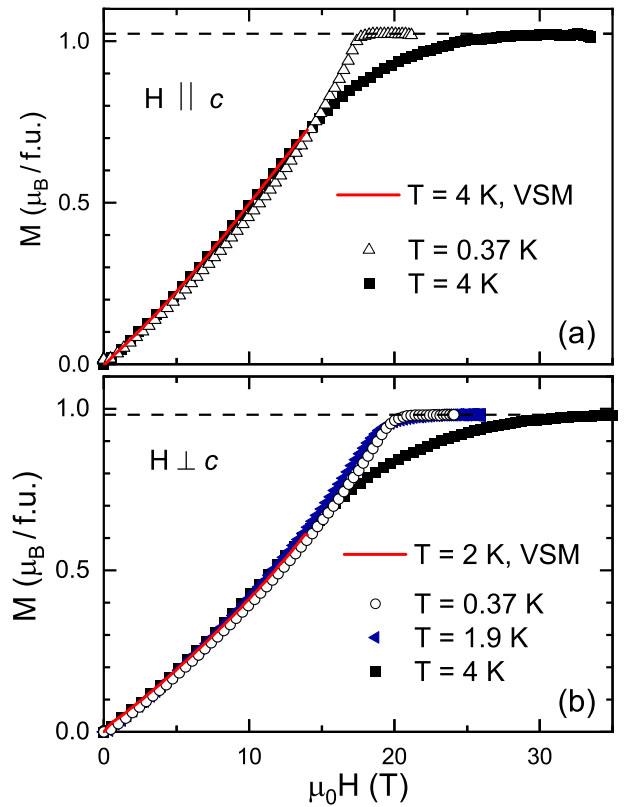

FIG. S3. Pulsed-field magnetization at various temperatures for (a) out-of-plane and (b) in-plane field directions and comparison to magnetization curves recorded with a commercial VSM magnetometer in DC fields up to 14 T.

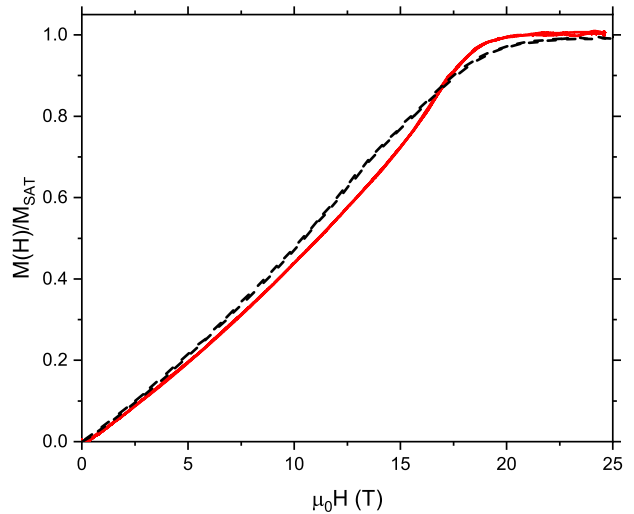

FIG. S4. High-field magnetization of a polycrystalline CuPOF sample at 0.63 (pulsed-fields, red solid line) and 1.3 K (DC fields, black dashed lines), recorded at the National High Magnetic Field Laboratory in the Los Alamos and Tallahassee facilities, respectively.

temperature ESR linewidth in the $ac$, $bc$, and $ab$ planes

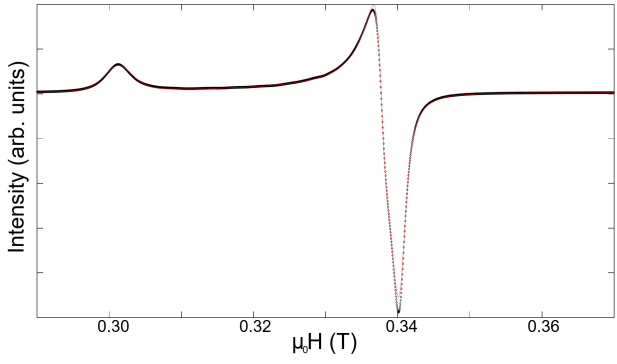

FIG. S5. ESR powder spectrum of CuPOF at room temperature measured at 9.8 GHz. The spectrum is shown as small black diamonds and the least squares fit to the data is shown as red line.

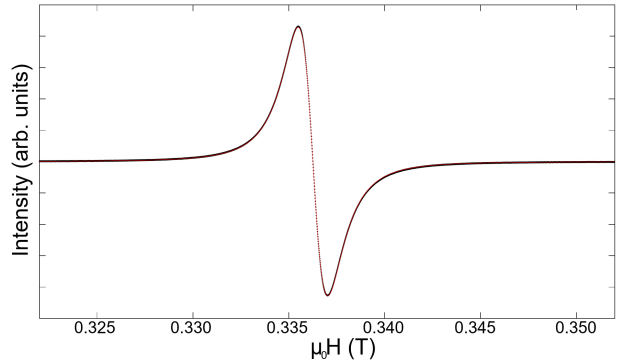

FIG. S6. ESR spectrum of $[Cu(pyz)_2(2\text{-HOpy})_2](PF_6)_2$ with field applied along the $a$ axis measured at 9.8 GHz. A fit with a pseudo-Voigt function yields a Lorentzian contribution of 1.15468 mT and a Gaussian contribution of 0.0051 mT.

TABLE S3. Comparison of the anisotropic $g$-factors and ESR linewidths, obtained from measurements of a polycrystalline sample (PCU), as well as single-crystalline samples at Clark University (CCU) and at the Dresden High Magnetic Field Laboratory (HLD). The average values $\langle g_{sc} \rangle$ and $\langle \Delta H_{sc} \rangle$ of the two independent measurements of the single-crystalline CuPOF samples are presented as well.

| $g$-factors | PCU | CCU | HLD | $\langle g_{sc} \rangle$ |
|---|---|---|---|---|
| $g_a$ | 2.071 | 2.074(1) | 2.072(1) | 2.073(2) |
| $g_b$ | 2.056 | 2.068(2) | 2.063(2) | 2.066(4) |
| $g_c$ | 2.322 | 2.300(1) | 2.296(1) | 2.298(2) |
| $\langle g \rangle$ | 2.150 | 2.147(1) | 2.144(1) | 2.146(3) |
| Linewidth (Oe) | | | | $\langle \Delta H_{sc} \rangle$ |
| $\Delta H_{pp}^a$ | | 15.7 | 16.0 | 15.9(2) |
| $\Delta H_{pp}^b$ | | 15.3 | 15.1 | 15.2(1) |
| $\Delta H_{pp}^c$ | | 21.7 | 21.3 | 21.5(2) |

are shown in Fig. S8, and were described by means of the

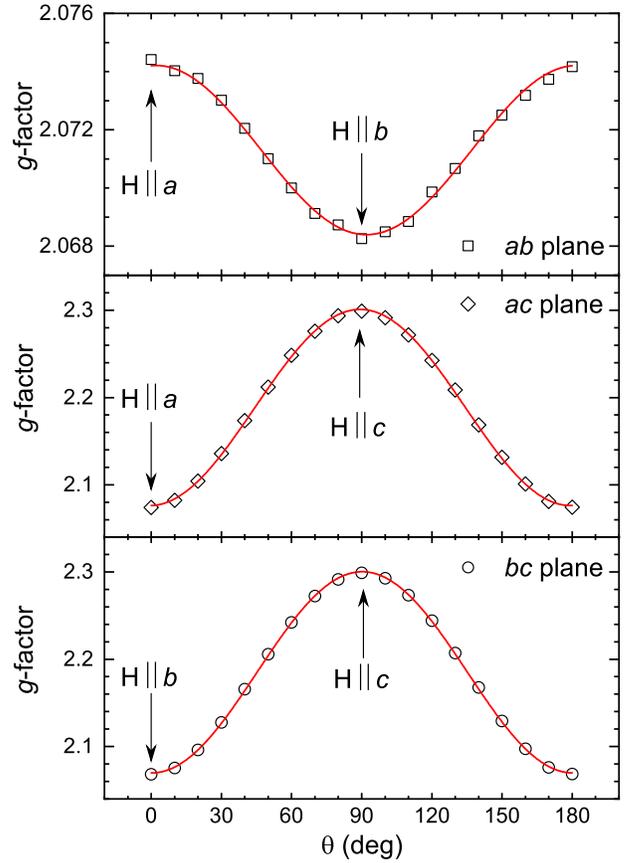

FIG. S7. Angular variation of the $g$-factor of $[Cu(pyz)_2(2\text{-HOpy})_2](PF_6)_2$ at room temperature measured at 9.8 GHz for three orthogonal planes. The anisotropies of the $g$-factor were modeled using a $\cos^2(\theta)$-type angular dependence, as represented by the red solid lines. Note the fine scale for the $ab$-plane variation.

Hamiltonian

$$\mathcal{H} = \mathcal{H}_Z + \mathcal{H}_J + \mathcal{H}_{D/d}, \quad (1)$$

where $\mathcal{H}_Z$ is the Zeeman interaction, $\mathcal{H}_J$ is the isotropic exchange interaction, and $D$ and $d$ are the anisotropic and antisymmetric exchange interactions, represented by the term $\mathcal{H}_{D/d}$. Due to the relatively strong isotropic exchange, $J/k_B = 6.80(5)$ K, the hyperfine interaction was neglected. The anisotropy of the $g$-factor likely accounts for some of the broadening observed in the $c$ direction. Based on the powder spectrum of CuPOF, a minor Gaussian broadening of about 0.267 Oe is expected from the anisotropy of the $g$-factor. Due to the large Cu–Cu distance, the dipolar interaction in the $c$ direction is expected to be small. A possible spin diffusive behavior, expected for 2D systems [3], is not observed. In the $ab$ plane, no angular variation of the linewidth is observed within experimental uncertainty. The angular variation of the linewidth in the $ac$ and $bc$ planes was interpreted as the sum of the isotropic scaling factor $A$ and the variation of the linewidth due to the antisymmetric and anisotropic

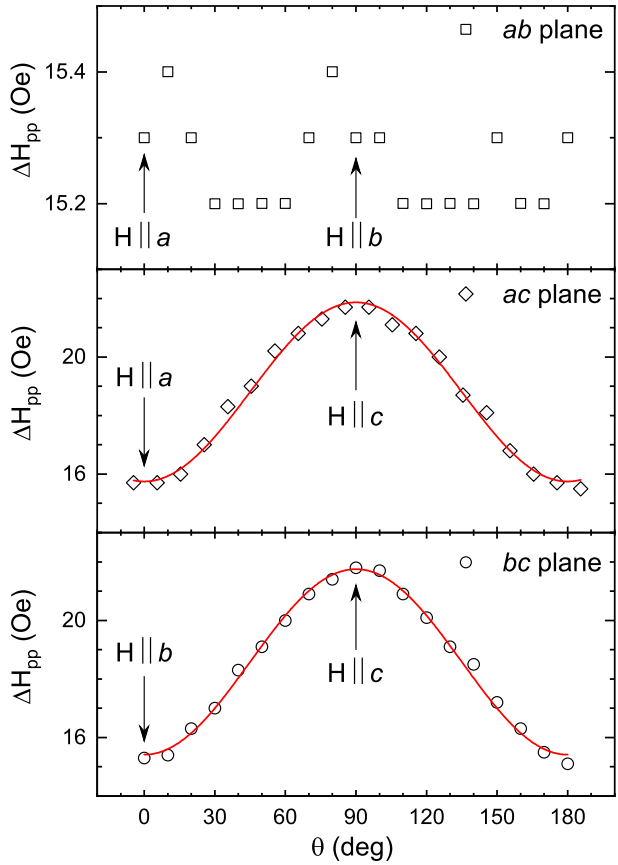

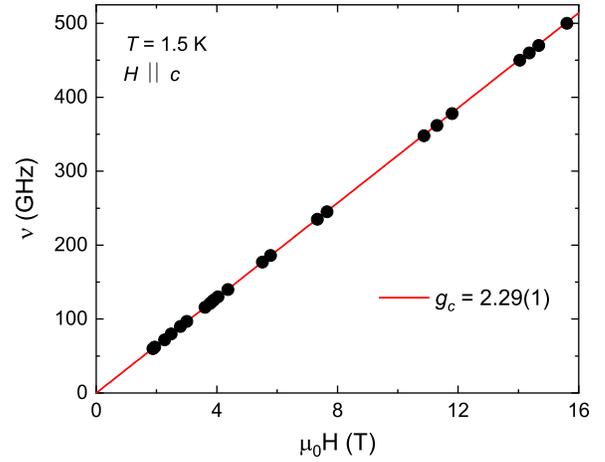

FIG. S9. The out-of-plane ESR frequency – magnetic field diagram at 1.5 K.

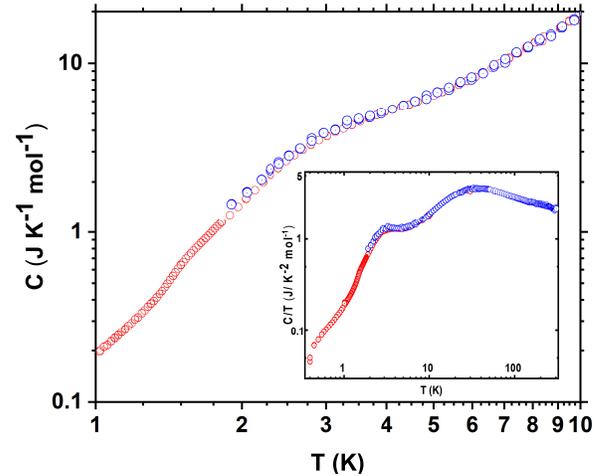

FIG. S10. Temperature dependence of the specific heat of [Cu(pyz)$_2$(2-HOpy)$_2$](PF$_6$)$_2$ in the temperature range between 1 and 10 K. The inset displays the same data in the temperature range between 0.4 and 300 K. The data recorded by use of the $^3$He and $^4$He cryostats are denoted by the red and blue circles, respectively.

FIG. S8. ESR peak to peak linewidth as a function of angle in the $ab$, $ac$, and $bc$ planes. The red solid lines represent fits to the data with equation (2). Note the fine scale for the $ab$-plane variation.

exchange $B$, yielding

$$\Delta H_{pp}(\theta) = A + B\cos^2\theta. \qquad (2)$$

The parameters found from modeling of Eq. (2) to the experimental data for the $ac$ ($bc$) plane, represented by the red lines in Figs. S8(b) and S8(c), are $A = 15.73(8)$ Oe [$A = 15.41(7)$ Oe] and $B = 6.17(14)$ Oe [$B = 6.34(12)$ Oe].

The ESR frequency–field diagram at 1.5 K for $H\parallel c$ is presented in Fig. S9. A single line corresponding to only one resonance mode was observed at all frequencies up to 500 GHz. The frequency-field dependence represents transitions between the Zeeman-driven splitting of energy levels of the Cu$^{2+}$ ions, and can be described by the linear relation $h\nu = g\mu_B B$ in the paramagnetic state. The low-temperature $g$-factor, which determines the slope of the frequency-field diagram, was obtained as $g_c = 2.29(1)$ from a linear fit of the experimental data, denoted by the red solid line in Fig. S9. Measurements at a temperature significantly lower than $T_N = 1.38(2)$ K would be required in order to detect a possible energy gap, and to explore the ESR excitation spectrum in the ordered state.

*Specific Heat.* The specific heat at temperatures between 0.4 and 300 K is presented in the inset of Fig. S10. The data recorded by use of $^3$He and $^4$He cryostats are denoted by the red and blue circles, respectively. No sharp anomalies corresponding to structural changes or ordering transitions were observed in this range.

The specific heat between 1 to 10 K reveals a broad hump exceeding the phonon contribution, see main panel of Fig. S10. The $^3$He data were analyzed as the sum of the magnetic specific heat of a 2D QHAF and a phononic contribution. The magnetic specific heat capacity was

TABLE S4. Coefficients for the magnetic specific heat of the square-lattice 2D QHAF.

| Index | $N_i$ | $D_i$ |
|-------|-------|-------|
| 1 | 0.00657 | 1.86131 |
| 2 | 0.00761 | -10.93035 |
| 3 | -0.16066 | 28.4599 |
| 4 | 2.913 | -32.816 |
| 5 | -0.35042 | 20.93204 |

represented as a ratio of polynomials

$$C_{mag} = R \frac{\sum_{i=1}^{5} N_i \, (T/J)^i}{\sum_{i=1}^{5} D_i \, (T/J)^i}, \qquad (3)$$

where $R$ is the gas constant and the values of the coefficients $N_i$ and $D_i$ are given in Table S4. This new polynomial is similar in form to one used in a previous study [4], but is based on recent quantum Monte Carlo simulations of the magnetic specific heat [5] that extended to lower relative temperatures and can also represent the specific heat of rectangular lattices in which the exchange strengths ($J$ and $\alpha J$) along the $a$ and $b$ axes are different. The range of validity for the square lattice is between $0.15 \leq T/J \leq 5.0$.

*Muon Spin Relaxation.* In a $\mu^+$SR experiment [6], spin-polarized positive muons are stopped in a target sample, where the muon usually occupies an interstitial position in the crystal. The observed property in the experiment is the time evolution of the muon spin polarization, the behavior of which depends on the local magnetic field at the muon site. Each muon decays, with an average lifetime of 2.2 $\mu$s, into two neutrinos and a positron, the latter particle being emitted preferentially along the instantaneous direction of the muon spin. Recording the time dependence of the positron emission directions, therefore, allows the determination of the spin polarization of the ensemble of muons. In our experiments positrons are detected by detectors placed forward (F) and backward (B) of the initial muon polarization direction. Histograms $N_F(t)$ and $N_B(t)$ record the number of positrons detected in the two detectors as a function of time following the muon implantation. The quantity of interest is the decay positron asymmetry function, defined as

$$A(t) = \frac{N_F(t) - \alpha_{exp} N_B(t)}{N_F(t) + \alpha_{exp} N_B(t)}, \qquad (4)$$

where $\alpha_{exp}$ is an experimental calibration constant. $A(t)$ is proportional to the spin polarization of the muon ensemble.

*Magic-Angle Spinning NMR spectroscopy.* $^{13}$C magic-angle spinning (MAS) NMR spectra were recorded employing a Bruker AVANCE-II-600 spectrometer in a 14.1 T magnetic field, using a home-built MAS probe for 4×25 mm $Si_3N_4$ rotors. 137 mg of $[Cu(pyz)_2(2-HOpy)_2](PF_6)_2$ powder was packed into the rotor and spun at 13.5 kHz. The temperature dependence of the spectrum was recorded between 276 and 332 K. Cooling of the sample was maintained using a gas cooling unit BCUX from Bruker. The actual temperature of the rotating sample was determined by $^{207}$Pb chemical shift of lead nitrate rotating under identical conditions [7]. The temperature variation within the sample was typically ±1.5 K, and ±3 K for temperatures above 300 K.

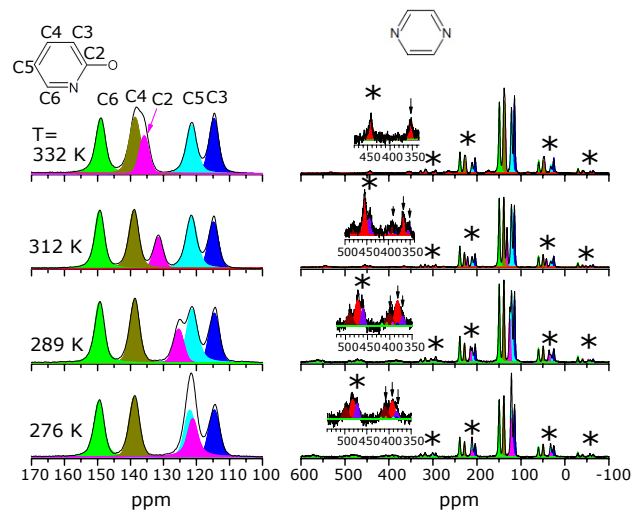

FIG. S11. Temperature dependence of the $^{13}$C MAS NMR spectrum of $[Cu(pyz)_2(2-HOpy)_2](PF_6)_2$. Full spectra are shown in the right panel. The sharp peaks in the region between 110 to 150 ppm correspond to the carbon sites in the 2-pyridone molecules, and are shown in detail in the left panel. The asterisks in the right panel denote spinning side lines at a distance of multiple of the spinning speed from the main lines. The large number of spinning sidebands indicates large anisotropies of the magnetic shift tensors. The weak lines in the region between 350 to 420 ppm denoted by arrows in the insets correspond to the carbon sites in the pyrazine molecules.

Room-temperature $^{31}$P spectra have been recorded on a Bruker AVANCE-II-360 spectrometer in an 8.45 T external magnetic field with a $^{31}$P resonance frequency of 145.56 MHz, using a home-built MAS probe for 1.8×15 mm rotors. Variable-temperature $^{31}$P MAS NMR spectra were studied at temperatures between 27 and 310 K. Low-temperature spectra have been recorded with the Bruker AVANCE-II spectrometer at an external field of 4.7 T with a $^{31}$P resonance frequency of 80.985 MHz, using a home-built cryoMAS probe [8] for 1.8 mm $Si_3N_4$ rotors. The sample spinning speed was set to 30 kHz at all temperatures except the lowest experiment at 27 K, where the spectrum was recorded at 25 kHz spinning rate. This slowing down of the spinning rate is necessary to stabilize such a low temperature. The temperature of the spinning sample was measured with a temperature sensor at the spinner assembly and corrected for a given spin-ning frequency with the temperature dependent $^{207}$Pb chemical shift of $Pb(NO_3)_2$ [7], which was rotating at the same conditions.

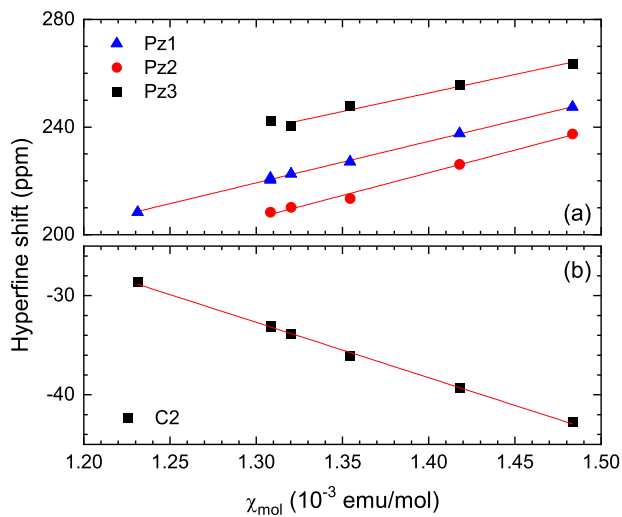

FIG. S12. Hyperfine shift of the pyrazine carbons (a) and the carbon C2 site of the 2-pyridone molecule (b) in CuPOF.

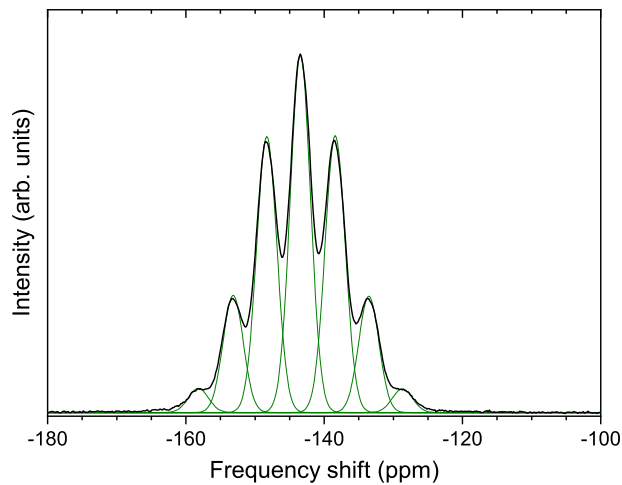

FIG. S13. Room-temperature $^{31}$P MAS NMR spectrum of $[Cu(pyz)_2(2\text{-}HOpy)_2](PF_6)_2$ at 145.56 MHz and a sample spinning frequency of 40 kHz.

The temperature dependence (276 – 332 K) of the $^{13}$C MAS NMR spectrum is shown in Fig. S11. All five resonances of the 2-pyridone molecules are resolved and appear at frequency shifts between 110 and 150 ppm, see left panel in Fig. S11. Spinning sidebands are also observed and marked by asterisks in the right panel of Fig. S11. The frequency shifts of the 2-pyridone carbons are close to the regular shift values and are independent of temperature, except for the carbon site C2. There are four independent carbons on the two independent pyrazine molecules. However, only three MAS NMR spectral lines have been resolved at the lowest temperature measured. These lines appear between 350 to 420 ppm with monotonous increase of the frequency shift with decreasing temperature, see Table S5. The hyperfine shift of the carbon sites can be evaluated as $K = \delta_{OBS} - \delta_{CS}$, where $\delta_{OBS}$ and $\delta_{CS}$ are the observed frequency shift and the chemical shift value in solution, respectively.

The hyperfine shift $K$ is caused by the magnetic moments of the copper $d$-shell electrons. It is related to the molar magnetic susceptibility $\chi_{mol}$ by $H_{hf}\chi_{mol}/N_A\mu_B$, where $H_{hf}$ is the hyperfine field, $\mu_B$ is the Bohr magneton and $N_A$ is Avogadro's number. The hyperfine field can be determined from the plot $K$ vs $\chi_{mol}$. Fig. S12 shows that the hyperfine shift of the 2-pyridone C2 carbon (b) and that of the pyrazine carbons (a) linearly depend on the magnetic susceptibility. The slopes in the plots give the positive hyperfine field, $H_{hf}$, values 770(60), 861(13), and 940(50) $Oe/\mu_B$ for the pyrazine carbon sites and the negative value -312(7) $Oe/\mu_B$, for the carbon site C2 of the 2-pyridone molecule.

The room-temperature $^{31}$P-MAS NMR spectrum is shown in Fig. S13. Due to the cancellation of local dipole fields by spinning the sample with a frequency of 40 kHz at the magic-angle orientation, the $^{31}$P-MAS spectrum is revealed as $J$ resolved with 7 lines, separated by a scalar spin-spin coupling of $J = 712$ Hz between the $^{19}$F and $^{31}$P nuclei. The chemical shift of -143.2 ppm and $J$ are in good agreement with previously reported values for compounds containing $PF_6$ molecules [9, 10]. As the hyperfine fields at the phosphorus sites and chemical shift variation of the PF6 anions are small, the two phosphorus sites are not resolved in the MAS experiments. The absence of considerable spinning sidebands of the resonance indicates an isotropic surrounding of the phosphorus.

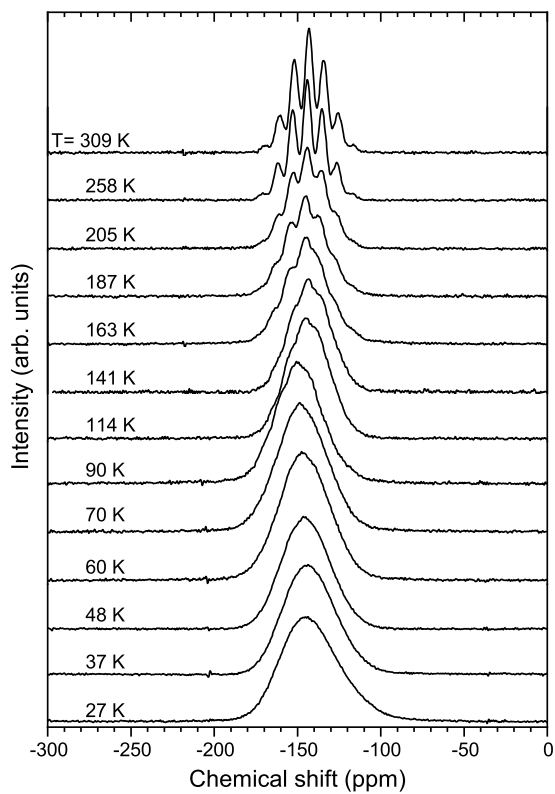

FIG. S14. Temperature dependence of the $^{31}$P MAS NMR spectrum of $[Cu(pyz)_2(2\text{-}HOpy)_2](PF_6)_2$ at 80.985 MHz.

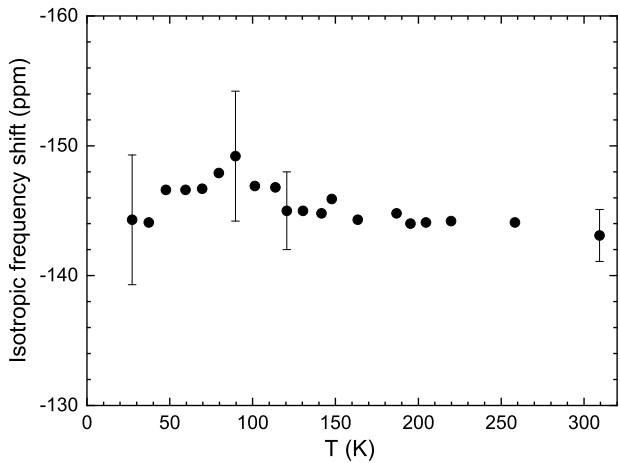

FIG. S15. Temperature dependence of the $^{31}$P isotropic value of frequency shift at 80.985 MHz. $K_{ISO}$ was determined as the frequency of the peak maximum.

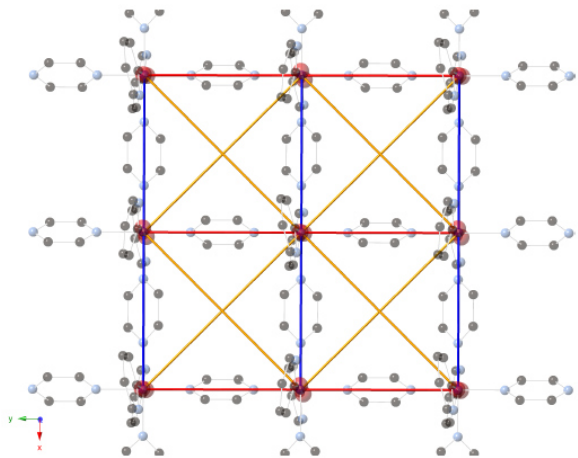

FIG. S16. Color-code for the relevant magnetic interactions computed within a 2D layer, whose molecular structure is here de-emphasized for clarity. Color code: $J_1$(red), $J_2$(blue), and $J_3$ (orange). $J_4$ accounts for the interlayer coupling (perpendicular to the plane, not shown).

The temperature-dependent $^{31}$P-MAS NMR spectra are shown in Fig. S14. At high temperatures, the $^{31}$P-MAS NMR spectrum appears $J$ resolved with 7 components, as described above for the room-temperature spectrum. Towards lower temperatures, the components become broader and the resonance becomes gradually a single slightly asymmetric line. Usually, the main broadening mechanism in paramagnetic rotating solids is the influence of the magnetic susceptibility of the powder particles [11].

TABLE S5. Frequency shift of the carbon sites in the 2-pyridone molecule (noted as C2-C6) and in the pyrazine molecules (Pz1, Pz2, Pz3). The chemical shifts (CS) in solution are given in the last row.

| $T$ (K) | C6 | C4 | C2 | C5 | C3 | Pz1 | Pz2 | Pz3 |
|---|---|---|---|---|---|---|---|---|
| 276 | 149.4 | 138.6 | 121.8 | 120.9 | 114.3 | 392.4 | 382.5 | 408.6 |
| 288.5 | 149.3 | 138.5 | 125.2 | 121.0 | 114.4 | 382.7 | 371.1 | 400.6 |
| 301.8 | 149.1 | 138.5 | 128.4 | 121.2 | 114.3 | 372.1 | 358.4 | 392.7 |
| 312.3 | 149.2 | 138.8 | 131.4 | 121.4 | 114.7 | 366.1 | 353.3 | 366.1 |
| 331 | 149.1 | 138.7 | 135.9 | 121.4 | 114.8 | 353.4 | | |
| CS | 142.8 | 136.6 | 164.5 | 119.0 | 107.0 | 145.0 | | |

The frequency shift of the $^{31}$P resonance line shows almost no variation with temperature, see Fig. S15. Usually, the paramagnetic shift is proportional to the magnetic susceptibility. With decreasing temperature from 300 to 27 K the latter increases significantly (see Fig. 2 in main text), whereas there is nearly no variation of the isotropic value of the $^{31}$P NMR frequency shift in that temperature range.

*Density Functional Theory (DFT) Calculations.* The magnetic exchange couplings $J$ between nearest-neighbor (NN) and next-nearest-neighbor (NNN) copper pairs were analyzed using the isotropic Heisenberg Hamiltonian

$$\mathcal{H} = \sum_{A,B}^{N} J_{AB} \mathbf{S}_A \cdot \mathbf{S}_B. \quad (5)$$

The broken-symmetry (BS) approach [12, 13] has been used to properly describe the open-shell low-spin states (LS, BS), while the equation proposed by Yamaguchi and co-workers [14] has been employed to account for the different weight of the open-shell and closed-shell states in the BS solution. All calculations have been performed with GAUSSIAN 09 [15] at the UB3LYP [16–18] level, and using a triple-zeta polarization basis set (TZVP).

The magnetic exchange couplings ($J_1$, $J_2$, $J_3$, and $J_4$) have been computed for four possible types of dimers in the unit cell of $[Cu(pyz)_2(2-HOpy)_2](PF_6)_2$. Pairs separated by $d_1$ (6.676 Å) and $d_2$ (6.680 Å) correspond to the NN interaction through the pyrazine molecules, while $d_3$ (9.727 Å) is the diagonal next-nearest-neighbor interaction within the magnetic layer. In turn, $d_4$ (13.10 Å) is the nearest-neighbor distance between Cu units in adjacent layers. Pairs at larger distances have not been considered. The cluster models selected to compute $J_1$ to $J_3$ have been designed to incorporate the potentially-relevant counter ions in an explicit manner. Ideally, the treatment of the whole unit cell of the crystal, including periodic boundary conditions, would be necessary to properly account for effects of the environment in the $J$ values. However, the small magnitude of the magnetic interactions make them incompatible with the approxi-mations adopted in this type of calculations to keep the computational cost under control. In the present case, with the aim at balancing cost and accuracy, we have used a tetramer cluster model, i.e., four Cu centers, for the $J_1$ to $J_3$ intralayer interactions, rather than the common dimer model [19]. For comparison purposes, the dimer model has also been used to compute $J_1$ to $J_3$ and the interlayer value $J_4$. The cluster models are shown in Fig. S17.

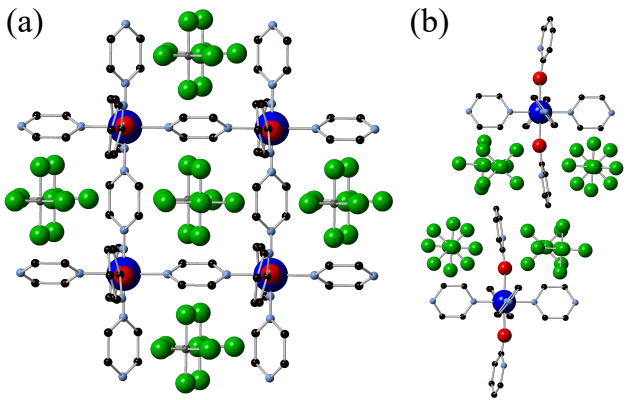

(a)

(b)

FIG. S17. Cluster models employed for the calculation of $J_1$ to $J_4$. (a) Tetramer model accounting for the intralayer interactions $J_1$, $J_2$, and $J_3$. Note that $PF_6$ molecules are located above and below the plane formed by the Cu-pyrazine network. (b) Dimer model used to describe the shortest interlayer interaction $J_4$. Color code for atoms: copper (blue, highlighted), oxygen (red), nitrogen (pale blue), phosphorus (gray), fluorine (green), carbon (black). Hydrogen atoms are not shown for clarity.

TABLE S6. Magnetic interactions $J_1/k_B$ to $J_4/k_B$ and associated Cu–Cu distances $d_1$ to $d_4$ computed using the various cluster models. CI = Counter Ions.

| $J_i/k_B$ | $d_i$ (Å) | Tetramer+CI (K) | Dimer+CI (K) | Dimer (K) |
|---|---|---|---|---|
| $J_1/k_B$ | 6.676 | 18.7 | 29.3 | 11.8 |
| $J_2/k_B$ | 6.680 | 16.4 | 29.3 | 11.8 |
| $J_3/k_B$ | 9.727 | 0.3 | $< |0.15|$ | $< |0.15|$ |
| $J_4/k_B$ | 13.097 | $< |0.15|$ | $< |0.15|$ | $< |0.15|$ |

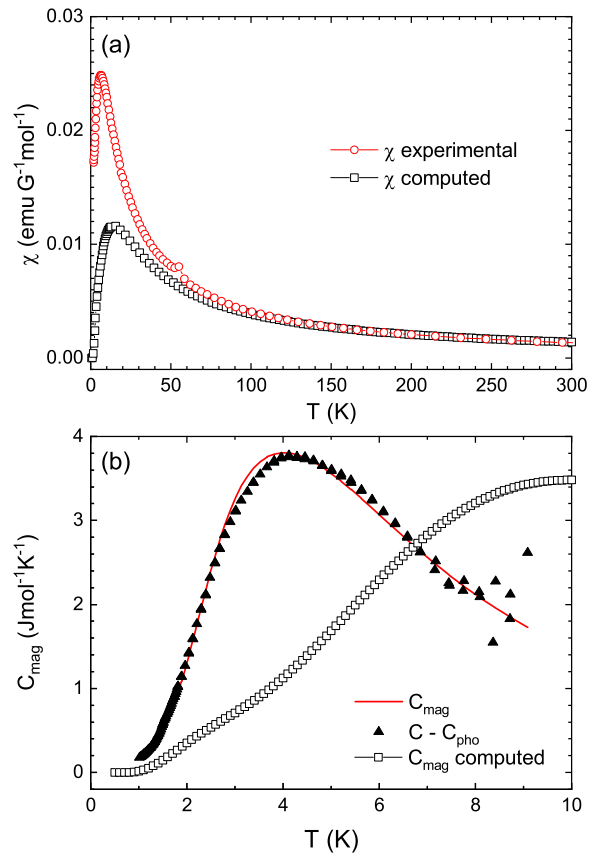

FIG. S18. Computed and experimentally-measured (a) magnetic susceptibility and (b) specific heat for $[Cu(pyz)_2(2\text{-}HOpy)_2](PF_6)_2$.

The computed magnetic interactions confirm the 2D magnetic topology of the system, see Table S6. The dominant magnetic interactions are between adjacent copper centers in the layers, with $J_1/k_B$ of 18.7 K and $J_2/k_B$ of 16.4 K. The diagonal interaction, $J_3/k_B$, is much weaker with only 0.3 K, whereas the interlayer interaction is negligible, with $J_4/k_B < |0.15|$ K. Notice that the strengths of $J_1$ and $J_2$ are slightly different despite the near equality of the respective copper-copper distances. Such a difference was also observed in copper pyrazine perchlorate $Cu(pz)_2(ClO_4)_2$ [20] and was ascribed to (i) the different relative disposition of the counterions and (ii) to a change in the Cu–N and N–N distances involved in the magnetic pathways.

The calculations using the dimer cluster model yield $J_1/k_B = J_2/k_B = 29.3$ K, and $J_3/k_B = J_4/k_B < |0.15|$ K. Note that $J_1$ and $J_2$ are equivalent within this model, and $J_3$ vanishes in comparison with the calculations using a tetramer model. Finally, the dimer cluster model in the absence of counter ions leads to the weakening of $J_1/k_B = J_2/k_B = 11.8$ K. This suggests, similar to the case of $Cu(pz)_2(ClO_4)_2$ [20], that the counterions have a strong influence on the magnetic pathway, enhancing the strength of the antiferromagnetic interactions. It is conceivable that further enlargement of our cluster model, by increasing either the number of Cu centers or the neighboring $PF_6$ molecules, would change our computational estimation of $J_1$. Generally, the DFT calculations confirm the magnetic topology of $[Cu(pyz)_2(2\text{-}HOpy)_2](PF_6)_2$, as consisting of magnetically isolated 2D layers.

Using the computational estimates of $J_1 - J_3$ by the tetramer cluster model, the macroscopic properties of $[Cu(pyz)_2(2\text{-}HOpy)_2](PF_6)_2$ can be computed by using well-known statistical-thermodynamics expressions within the first-principles bottom-up procedure (FPBU) [21, 22]. A subset of the magnetic topology, called *model space*, is selected in a way that, when the Heisenberg Hamiltonian is applied on the basis of a regionally reduced density-matrix approach, the resulting set of eigenvalues reproduces those that result from the application of the Heisenberg Hamiltonian to the full, infinite crystal. In the present study, we have selected a model space of sixteen Cu atoms arranged in a $4 \times 4$ plane, which aims at reproducing the infinite 2D layers of $[Cu(pyz)_2(2\text{-}HOpy)_2](PF_6)_2$. Sixteen sites are usually sufficient to achieve convergence of the magnetic properties, especially considering the simplicity of the magnetic topology of $[Cu(pyz)_2(2\text{-}HOpy)_2](PF_6)_2$. Finally, the matrix representation of the Heisenberg Hamiltonian is built and fully diagonalized by using as a basis set the space of $m_s$ spin functions of a magnetic model of the selected 16 sites. The resulting energy and spin multiplicity of all possible magnetic states (up to 12870) are then used to calculate the macroscopic properties of the system, such as the magnetic susceptibility and specific heat. As shown in Fig. S18, the experimental and calculated values are qualitatively similar, with temperature dependences that scale with the values of the interaction strengths.

[1] S. Stoll and A. Schweiger, J. Magn. Reson. **178**, 42 (2006).

[2] A. Bencini and D. Gatteschi, *EPR of Exchange Coupled Systems* (Dover Publications, 2012).

[3] P. M. Richards and M. B. Salamon, Phys. Rev. B **9**, 32 (1974).

[4] T. Lancaster, S. J. Blundell, M. L. Brooks, P. J. Baker, F. L. Pratt, J. L. Manson, M. M. Conner, F. Xiao, C. P. Landee, F. A. Chaves, S. Soriano, M. A. Novak, T. P. Papageorgiou, A. D. Bianchi, T. Herrmannsdörfer, J. Wosnitza, and J. A. Schlueter, Phys. Rev. B **75**, 094421 (2007).

[5] A. Sandvik and C. P. Landee, unpublished.

[6] S. J. Blundell, Contemp. Phys. **40**, 175 (1999).

[7] L. C. M. Gorkom, J. M. Hook, M. B. Logan, J. V. Hanna, and R. E. Wasylishen, Magn. Reson. Chem. **33**, 791 (1995).

[8] A. Samoson, T. Tuherm, J. Past, A. Reinhold, T. Anupold, and I. Heinmaa, Top. Curr. Chem. **246**, 15 (2005).

[9] E. R. Andrew, M. Firth, A. Jasinski, and P. Randall, Phys. Lett. A **31**, 446 (1970).

[10] E. C. Alyea, J. Malito, and J. H. Nelson, Inorg. Chem. **26**, 4294 (1987).

[11] M. Alla and E. Lippmaa, Chem. Phys. Lett. **87** (1982).

[12] L. Noodleman and E. R. Davidson, Chem. Phys. **109**, 131 (1986).

[13] L. Noodleman, J. Chem. Phys. **74**, 5737 (1981).

[14] T. Soda, Y. Kitagawa, T. Onishi, Y. Takano, Y. Shigeta, H. Nagao, Y. Yoshioka, and K. Yamaguchi, Chem. Phys. Lett. **319**, 223 (2000).

[15] M. J. Frisch, G. W. Trucks, H. B. Schlegel, G. E. Scuseria, M. A. Robb, J. R. Cheeseman, G. Scalmani, V. Barone, B. Mennucci, G. A. Petersson, H. Nakatsuji, M. Caricato, X. Li, H. P. Hratchian, A. F. Izmaylov, J. Bloino, G. Zheng, J. L. Sonnenberg, M. Hada, M. Ehara, K. Toyota, and Fukuda, Inc. **4** (2009).

[16] A. D. Becke, J. Chem. Phys. **104**, 1040 (1996).

[17] A. D. Becke, Phys. Rev. A **38**, 3098 (1988).

[18] C. Lee, W. Yang, and R. G. Parr, Phys. Rev. B **37**, 785 (1988).

[19] S. Vela, A. Sopena, J. Ribas-Arino, J. J. Novoa, and M. Deumal, Chem. Eur. J. **20**, 7083 (2014).

[20] S. Vela, J. Jornet-Somoza, M. M. Turnbull, R. Feyerherm, J. J. Novoa, and M. Deumal, Inorg. Chem. **52**, 12923 (2013).

[21] J. J. Novoa, M. Deumal, and J. Jornet-Somoza, Chem. Soc. Rev. **40**, 3182 (2011).

[22] M. Deumal, M. J. Bearpark, J. J. Novoa, and M. A. Robb, J. Phys. Chem. A **106**, 1299 (2002).